\documentclass[structabstract]{aa}  
\usepackage{graphicx}
\usepackage{natbib}
\usepackage{subfigure}
\bibpunct{(}{)}{;}{a}{}{,} 
\usepackage{txfonts}

\begin{document}

\def\Msun{\hbox{M$_{\odot}$}}
\def\Lsun{\hbox{L$_{\odot}$}}
\def\kms{km~s$^{\rm -1}$}
\def\hcop{HCO$^{+}$}
\def\n2hp{N$_{2}$H$^{+}$}
\def\micron{$\mu$m}
\def\13CO{$^{13}$CO}
\def\etamb{$\eta_{\rm mb}$}
\def\Inu{I$_{\nu}$}
\def\kapnu{$\kappa _{\nu}$}
\def\ffore{f$_{\rm{fore}}$}
\def\tastar{T$_{A}^{*}$}
\def\NH2{N(H$_{2}$)}
\def\deg{$^{\circ}$~}

\title{Characterizing Precursors to Stellar Clusters with Herschel
\thanks{{\it Herschel} in an ESA space observatory with science
instruments provided by European-led Principal Investigator
consortia and with important participation by NASA.}}
   \author{C. Battersby
     \inst{1}
     \and
     J. Bally\inst{1}
     \and
     A. Ginsburg\inst{1}
     \and
     J.-P. Bernard\inst{2,3}
     \and
     C. Brunt\inst{4}
     \and
     G. A. Fuller\inst{5}
     \and
     P. Martin\inst{6}
     \and
     S. Molinari\inst{7}
     \and
     J. Mottram\inst{4}
     \and
     N. Peretto\inst{5,8}
     \and
     L. Testi\inst{9,10}
     \and
     M. A. Thompson\inst{11}	  
   }
   \institute{Center for Astrophysics and Space Astronomy, University 
     of Colorado, Boulder, CO 80309, USA
     \and
     Universit\'{e} de Toulouse (UPS-OMP), Institut de Recherche en
     Astrophysique et Plan\'{e}tologie
     \and
     CNRS, UMR 5277, 9 Av. colonel Roche, BP 44346, F 31028 Toulouse cedex
     4, France 
     \and
     School of Physics, University of Exeter, Stocker Road, Exeter, EX4
     4QL, UK
     \and
     Jodrell Bank Centre for Astrophysics, School of Physics and
     Astronomy, University of Manchester, Manchester, M13 9PL, UK
     \and
     Canadian Institute for Theoretical Astrophysics, University of
     Toronto, Toronto, ON M5S 3H8, Canada
     \and
     INAF-Instituto, Fisica Spazio Interplanetario, via Fosso del
     Cavaliere 100, 00133 Roma, Italy
     \and
     Laboratoire AIM, CEA/DSM - INSU/CNRS - Universit\'{e} Paris Diderot,
     IRFU/SAp CEA-Saclay, 91191 Gif-sur-Yvette, France
     \and
     INAF - Osservatorio Astrofisico di Arcetri, Firenze, Italy
     \and
     European Southern Observatory, Garching bei Muenchen, Germany
     \and
     Centre for Astrophysics Research, Science and Technology Research
     Institute, University of Hertfordshire, Hatfield, UK
}



  \abstract
   {Despite their profound effect on the universe, the formation of
   massive stars and stellar clusters remains elusive.  Recent advances in
   observing facilities and computing power have brought us closer to
   understanding this formation process.  In the past decade, compelling
   evidence has emerged that suggests Infrared Dark Clouds (IRDCs) may be
   precursors to stellar clusters.  
   However, the usual method for identifying IRDCs is biased by the
  requirement that they are seen in absorption against background mid-IR
  emission, whereas dust continuum observations allow cold, dense
  pre-stellar-clusters to be identified anywhere.
   }
   {We aim to understand what physical properties characterize IRDCs, to
  explore the population of dust continuum sources that are not mid-IR-dark,
  and to roughly characterize the level of star formation activity 
  in dust continuum sources.}
   {We use Hi-GAL 70 to 500 \micron~data to identify dust continuum
  sources in the $\ell$=30\deg and $\ell$=59\deg Hi-GAL Science Demonstration Phase (SDP) 
  fields, to characterize and 
  subtract the Galactic cirrus emission, and
  perform pixel-by-pixel modified blackbody fits on cirrus-subtracted
  Hi-GAL sources.  We utilize archival Spitzer data to indicate the level of
  star-forming activity in each pixel, from mid-IR-dark to mid-IR-bright.}
   {We present temperature and column density maps in the Hi-GAL $\ell$=30\deg and
  $\ell$=59\deg SDP fields, as well as a robust algorithm for cirrus
  subtraction and source identification using Hi-GAL data. 
  We report on the fraction of Hi-GAL source pixels which are 
  mid-IR-dark, mid-IR-neutral, or mid-IR-bright in both fields.
  We find significant trends in column density and temperature
  between mid-IR-dark and mid-IR-bright pixels; mid-IR-dark pixels are about 10 K
  colder and have a factor of 2 higher column density on average than mid-IR-bright
  pixels.  We find that Hi-GAL dust continuum sources span a range of
  evolutionary states from pre- to star-forming, and that warmer sources
  are associated with more star formation tracers.  
  Additionally, there is a
  trend of increasing temperature with tracer type from mid-IR-dark at the coldest, to
  outflow/maser sources in the middle,
  and finally to 8 and 24 \micron~bright sources at the warmest.  Finally,
  we identify five candidate IRDC-like sources on the far-side of the Galaxy.
  These are cold ($\sim$ 20 K), high column
  density (N(H$_{2}$) $> 10^{22} \rm{cm}^{-2}$) clouds identified with Hi-GAL which, despite
  bright surrounding mid-IR emission, show little to no absorption at 8
  \micron.  These are the first inner Galaxy far-side candidate IRDCs
  of which the authors are aware.}  
   {}


   \maketitle
%

\section{Introduction}
Massive stars play a dominant role in shaping the Universe through their
immense ionizing radiation, winds, and spectacular explosive death, yet
their formation mechanism remains poorly understood.  
The dominant mode of star formation, and perhaps the only mode for massive
star formation seems to be clustered \citep{lad03, dew05}.  The definition
of clustered can be called into question \citep[e.g.][]{bre11, gie11},
but the search for young, massive star forming regions is still directed
toward `proto-clusters:' cold, dense, massive molecular clumps. 
Direct observation of proto-clusters is complicated: they are
rare, and therefore further away on average than isolated low mass
star-forming regions, proto-clusters have high column densities, meaning
that the proto-stars are highly embedded, and 
massive stars evolve rapidly and quickly heat and ionize their
surroundings, disrupting their natal molecular cloud.

Searches for proto-clusters or massive proto-stars are usually targeted at
long-wavelengths, as dust continuum emission from
these cold, dense sources peaks in the far-IR to sub-mm.
Surveys of molecular lines are another approach for identifying potential
proto-clusters, although these surveys may be time-intensive and the
strength of a molecular line is complicated by excitation conditions and
optical depth.
The discovery of Infrared Dark Clouds \citep[IRDCs][]{ega98,
per96,omo03}, opened a new window to viewing these cold, dense potential
proto-clusters in silhouette against the bright Galactic mid-IR background.  In the
past decade, compelling evidence has emerged that suggests that some IRDCs
may be precursors to massive stars and clusters \citep[e.g.][]{rat06, rag06,
beu07b, par09, bat10}.  
While there exist many small IRDCs \citep[e.g.][]{per10b, kau10}, the most
massive ones \citep[M $\sim 10^{3-4}$ \Msun,
n$_{H} > 10^{5} \rm{cm}^{-3}$, T $<$ 25 K,][]{rat06, ega98, car98} are
consistent with expectations for a proto-cluster.
However, the identification of an IRDC
requires that it be on the near-side of a bright mid-IR
background.  This limits our potential to understand the Galactic
distribution of potential proto-clusters.

Far-IR and submm dust continuum surveys are a powerful way to
identify proto-clusters throughout the Galaxy, as the cold, dense dust is optically
thin at these wavelengths.  Surveys such as the Bolocam Galactic Plane Survey
\citep[BGPS,][]{agu10} at 1.1 mm, the APEX Telescope Large Area Survey of
the Galaxy \citep[ATLASGAL,][]{sch09} at 870 \micron, and now Hi-GAL
\citep[The Herschel Infrared Galactic Plane Survey,][]{mol10a} from 70 to 500
\micron~are promising tools for
understanding star cluster formation on a Galactic scale.  However, the
sources identified in these surveys may span a large range of evolutionary
states, from pre-star-forming to star-forming, and further analysis or intercomparison
may be necessary to identify the pre-star-forming regions.  Since dust
temperatures and column densities can be derived from the multi-wavelength
Hi-GAL data, this data set allows for the distinction to be made between
pre- and star-forming regions.

For this study, we utilize data from the Hi-GAL survey \citep[][]{mol10a}
from 70 to 500 \micron~in the $\ell$=30\deg and
$\ell$=59\deg SDP fields to characterize dust continuum sources.  We investigate
differences in the physical properties of mid-IR-dark and mid-IR-bright clouds
identified in the dust continuum, and also the physical properties of
sources associated with various star formation tracers.
We use Extended Green Objects
\citep[EGOs, also known as ``green fuzzies'',][]{cyg08, cha09} 
to trace outflows from young stars, CH$_{3}$OH masers to trace 
sites of massive star formation, and 8 and 24
\micron~emission to indicate an accreting proto-star or UCHII Region
\citep{bat10}. 

This paper is organized as follows.  In Section
\ref{sec:obs} we introduce the Hi-GAL observing strategy and discuss
archival data used in our analysis.  In Section \ref{sec:meth} we describe the
Galactic cirrus emission removal and source identification methods, the modified
blackbody fitting procedure, and how the star formation tracers were
incorporated.  Section \ref{sec:res} describes our results, including a discussion
of uncertainties, the properties of the cirrus cloud emission, the
temperature and column density maps, the association of Hi-GAL sources with
mid-IR-dark and mid-IR-bright sources, and star formation tracers in Hi-GAL
sources.  In Section \ref{sec:farirdc} we present five candidate IRDC-like 
clouds on the far-side of the Galaxy, and finally, in Section
\ref{sec:conc} we summarize our conclusions.

   \begin{figure*}
   \centering
   \includegraphics[scale=0.8]{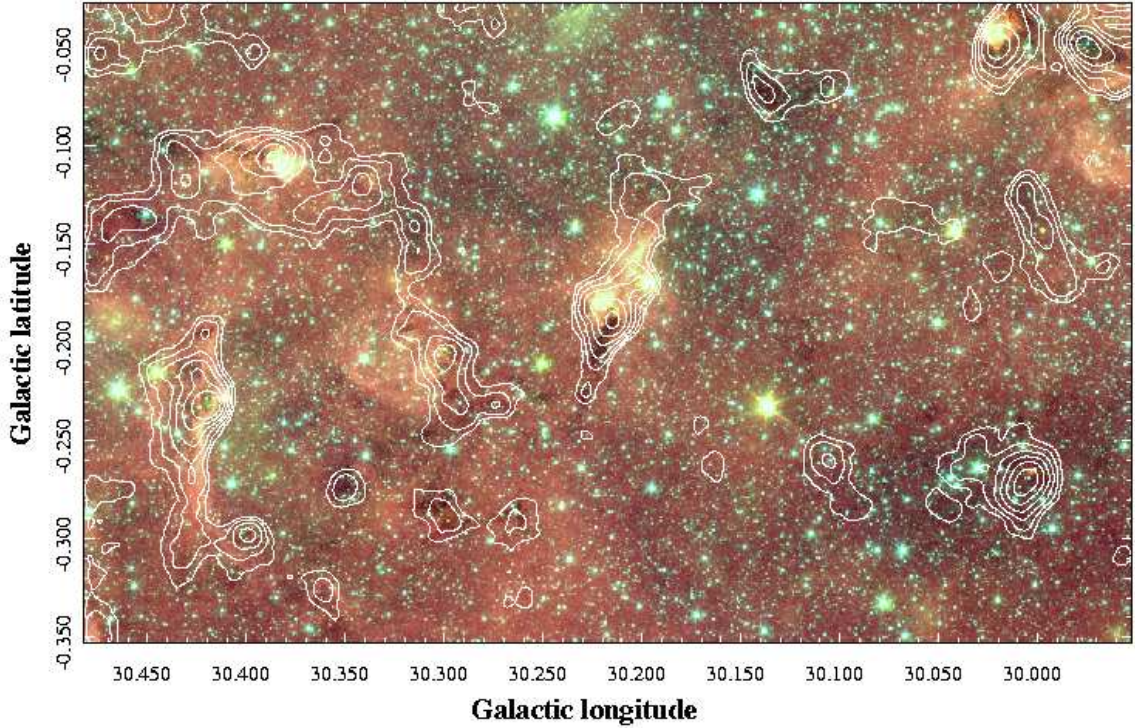}
   \caption{A three-color GLIMPSE image with logarithmic BGPS 1.1 mm dust continuum
   contours on our test region in the Hi-GAL SDP $\ell$=30\deg field.  
   Red: 8 \micron~, Green: 4.5 \micron~, and Blue: 3.6 \micron.
   This figure demonstrates the distinction between the IRDC and
   dust continuum population.  Some dust continuum sources are
   mid-IR-bright, some are mid-IR-dark, and some have no IR correlation.  The
   object in the center is particularly well-suited as a test case for this
   study as it contains both mid-IR-bright and mid-IR-dark dust continuum sources
   radiating as filaments on either side of a young HII region complex.}
   \label{fig:testirdc_glmrgb}
    \end{figure*}

\section{Observations and Archival Data}
\label{sec:obs}

\subsection{Hi-GAL}
The Herschel Infrared Galactic Plane Survey, Hi-GAL \citep{mol10a},
is an Open Time Key Project of the Herschel Space Observatory
\citep{pil10}.  Hi-GAL will perform a 5-band photometric survey of the
Galactic Plane in a $|b| \leq$ 
1\deg -wide strip from -70$^{\circ}$ $\leq$ l $\leq$ 70\deg at 70, 160,
250, 350, and 500 \micron~using the PACS \citep{pog10} and SPIRE
\citep{gri10} imaging cameras in parallel mode.  Two 2\deg $\times$ 2\deg
regions of the Hi-GAL survey were completed during the Science
Demonstration Phase \citep[SDP;][]{mol10a}, centered at approximately [$\ell$,b] = [30$^{\circ}$,
0$^{\circ}$] and [59$^{\circ}$, 0$^{\circ}$].

Data reduction was carried out using the \textit{Herschel} Interactive
Processing Environment \citep[HIPE,][]{ott10} with custom reduction scripts
that deviated considerably from the standard processing for PACS
\citep{pog10}, and to a lesser extent for SPIRE \citep{gri10}.  A
detailed description of the entire data reduction procedure can be found in
\citet{tra10}.  

The zero-level offsets in the Herschel maps were established by comparison 
with the IRAS and Planck data at comparable wavelengths, following the same 
procedure as described in \citet{Ber10}.
We compared the Herschel-SPIRE and PACS data with the predictions of a
model provided by the Planck collaboration (Planck Core-Team, private 
communication) and constrained on the Planck and IRIS data 
\citep{Miv05}.  The model uses the all-sky dust temperature maps
derived from the IRAS 100 \micron~and the two highest Planck frequencies 
to infer the average radiation field intensity for each pixel at the 
common resolution of the Planck and IRAS resolution of 5'. The Dustem model 
\citep{Com10} with
the above value for the radiation field intensity was then used to predict the
expected brightness in the Herschel-SPIRE and PACS bands, using the nearest
available Planck or IRAS band for normalization and taking into account the
appropriate color correction in the Herschel filters. The predicted brightness
was correlated with the observed maps smoothed to the 5' resolution
over the region observed with Herschel and the offsets were derived from
the zero intercept of the correlation.
We estimate the accuracy of the offset determination to better than 5\%.

\subsection{Archival Data}
We utilize the wealth of archival data in the Galactic Plane for our analysis.
We use the mid-IR data taken as part of the Galactic Legacy
Infrared Mid-Plane Survey Extraordinaire 
\citep[GLIMPSE; 3.6, 4.5, 5.8, and 8.0 \micron,][]{ben03} and the
MIPSGAL survey \citep[24 \micron;][]{car09}.  We also make use of the IRDC 
catalog of \citet{per09}, the catalog of Extended Green Objects
\citep[EGOs]{cyg08}, and the 1.1 mm
data and catalog from the Bolocam Galactic Plane Survey 
\citep[BGPS;][]{agu10, ros10}.  The Multi-Array Galactic Plane Imaging
Survey \citep[MAGPIS;][]{hel06,whi05}, which provides comprehensive radio
continuum maps of the first Galactic quadrant at high resolution and
sensitivity, is also used in our analysis.

\section{Methods}
\label{sec:meth}

\subsection{Source Definition and Removal of the Galactic Cirrus}

The Hi-GAL data reveal a wealth of structure in the Galactic Plane, from
cirrus clouds \citep{mar10} to filaments and clumps, as discussed in \citet{mol10b}.
Figure \ref{fig:testirdc_glmrgb} demonstrates the complicated association
of mid-IR and dust continuum sources toward the Galactic Ring.
In
this paper, we explore the physical properties of the densest components of
the Galactic Plane; the potential precursors to massive stars and
clusters.  In order to properly characterize these dense objects, a
careful removal of the Galactic cirrus is required.  We have explored a
variety of methods for the removal of the cirrus emission and we 
briefly discuss the pros, cons, and systematic 
effects of the different subtraction methods.  

Our first step is to project all the data onto a common grid, with a common
resolution and common units for comparison.  We crop
the images to the useable science field, convert the data
to units of MJy/sr, Gaussian convolve to a common resolution (36 \arcsec~and 25
\arcsec~for the lower and higher resolution modified blackbody fits), and regrid to a
common grid with a reasonable pixel size (roughly 1/4 the $\Theta_{FWHM}$).
These images are used in the remainder of the analysis.
We include images throughout the paper of a ``test field,'' (approximately,
[$\ell$,b] = [29.95, -0.35] to [30.50, 0]) chosen for it's
diversity of mid-IR-bright and mid-IR-dark sources.  The analysis was
orginally run and optimized on this region and then expanded to the full
Hi-GAL fields.

The first method for determining the dense clump source masks
was to use Bolocat clump masks derived from the Bolocam 
Galactic Plane Survey \citep[BGPS][]{agu10, ros10} at 1.1 mm.  While these
masks are a robust tracer of the cold, dense gas, they only extend to 
$|b| \leq$ 0.5\deg in the majority of our
science field and the sensitivity in the $\ell$=59\deg field is too poor
to trace the clumps seen in Hi-GAL.  

The following methods all use 
the SPIRE 500 \micron~data to determine the
source masks, as the cirrus decreases towards longer wavelengths 
\citep{gau92} and \citet{per10} demonstrate that the SPIRE 500 
\micron~data is well-suited for cirrus/dense source distinction. 
The second method we tried was based on that of \citet{per10}, who found
that toward IRDCs, the SPIRE 500 \micron~flux distribution showed a clear
peak at low fluxes with a long tail toward higher fluxes.  While this
method works well when applied to a single IRDC, as it was used in
\citet{per10}, it is not robust enough to be used to create a full Galactic
cirrus emission map from high ($|b| \gtrsim 1$ \deg) to low ($|b| = 0$ \deg)
Galactic latitudes.  We then tried a third method to create the 
dense source masks using a simple contrast map (contrast = data -
background) of the SPIRE 500 \micron~data and applied a cutoff.  This method
improved over the previous, however, it 
suffered from large negative bowls, creating source masks that were much
smaller than the physical source sizes.  In the fourth method, we define 
the background first by a second-order
polynomial plane fit (along Galactic latitude) to the smoothed 500
\micron~image.  This plane fit to the Galactic cirrus was then subtracted
from the original data and a cutoff was applied (determined by eye) to
define the dense source masks.  This
was the first method to do a reasonable job of identifying sources in a
range of Galactic latitudes.
We found, however, that a
polynomial plane was not a very good approximation to the shape of the
Galactic cirrus emission, and that the fit was particularly poor at high
Galactic latitudes ($|b| \sim 0.8 - 1$ \deg).

\begin{figure*}
   \centering
   \includegraphics[scale=0.8]{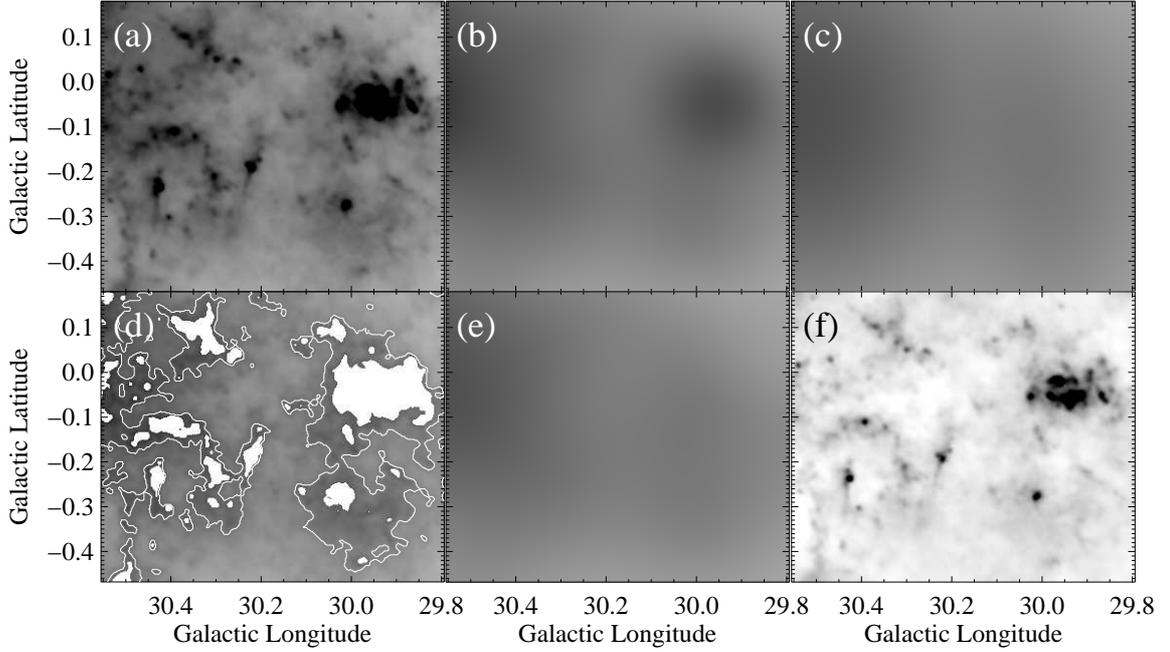}
   \caption{A depiction of the cirrus subtraction method for the first
   iteration. Panel (a) is
   the original SPIRE 500 \micron~image (all images on the same linear reverse
   grayscale, from -10 to 900 MJy/sr).  Panel (b) is the smoothed (FWHM
   Gaussian convolution kernel of 12') SPIRE 500 \micron~image that is
   used to fit the Gaussian
   shown in panel (c).  Panel (c) is then subtracted from panel
   (a) to produce a contrast image.  A 4.25 $\sigma$ cutoff is then applied to
   the contrast image to produce the source masks shown in panel (d).
   Panel (d) is the original SPIRE 500 \micron~image with the sources
   masked out in white.  This is the first iteration source masks; the
   final source boundaries are shown as white contours.  Panel (d) is then 
   convolved with a Gaussian FWHM
   convolution kernel of 12' with the masks treated as missing data to
   produce panel (e).  Panel (e) is considered the cirrus image for the
   next iteration.  In the next iteration, a Gaussian will be fit to the
   cirrus image, as a function of latitude at each longitude, and steps 
   b - e will repeat.
   Panel (f) is the difference image (original SPIRE 500 \micron~image -
   convolved cirrus image) that will eventually be used for 
   greybody fitting in the final iteration.}
   \label{fig:bgtest}
    \end{figure*}

\begin{figure}
     \centering
     \includegraphics[width=0.46\textwidth]{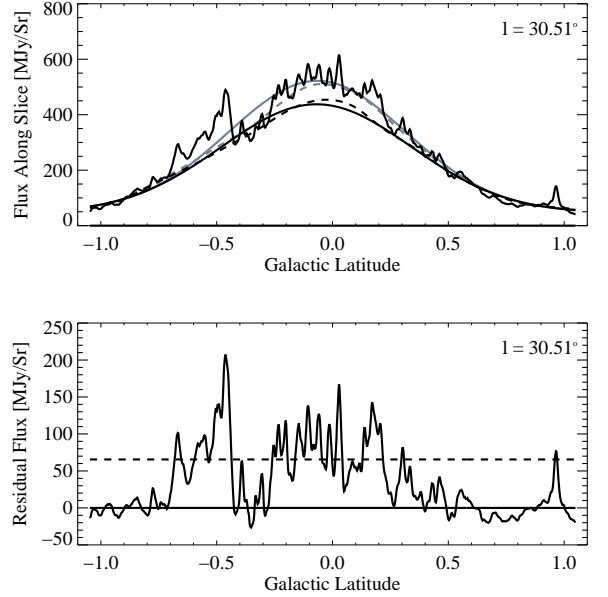}
     \caption{A slice across the Galactic Plane at l = 30.51\deg of the
     SPIRE 500 \micron~image, demonstrating the cirrus emission removal.  
     The top panel shows the SPIRE 500
     \micron~data, with a Gaussian fit to the smoothed background on the
     first iteration shown in a solid gray line.  The dashed gray line is
     the convolved background on the first iteration, after the source mask
     has been applied.  The black solid line is the Gaussian fit to the smoothed
     background on the final iteration and the black dashed line is the
     final convolved background after the source mask has been applied.
     Note that the final background (dashed black line) fits the low-lying
     diffuse emission nicely, including it's asymmetry about b =
     0\deg.  The bottom panel shows the final background subtracted science
     image cut along the same Galactic longitude, with the final source
     cutoff drawn as a dashed black line at 65 MJy/sr.  The success of the
     background subtraction method is demonstrated by the fact that the
     structure is flat across Galactic latitude and that the cutoff allows
     for the identification of the significant, though faint, source near b=1\deg.}
     \label{fig:bgcut}
   \end{figure}

Figure \ref{fig:bgtest} depicts the final method used to identify the dense
sources and to separate those from the cirrus cloud emission.  The first
panel (a) shows the original SPIRE 500 \micron~image. 
The first
step (shown in panel b) is to convolve the 500 \micron~image with a
Gaussian that is large enough to smooth over the sources but small enough
to capture variations in the cirrus emission.  We decided through trial and
error that a Gaussian with a FWHM of 12' was a reasonable compromise.  We then
fit a Gaussian in latitude to each Galactic longitude, as shown in Figure
\ref{fig:bgcut} and in panel (c) of Figure \ref{fig:bgtest}.  A Gaussian is
a reasonable approximation for the variation in the Galactic cirrus across
the Galactic Plane and worked to identify sources at high and low
Galactic latitudes equally well.  We then subtract the Gaussian fit
approximation of the Galactic cirrus from the original 500 \micron~data to
achieve a ``difference image.''  This ``difference image'' is the first
guess at the cirrus-subtracted source map.  

\begin{figure}
     \centering
     \includegraphics[width=0.46\textwidth]{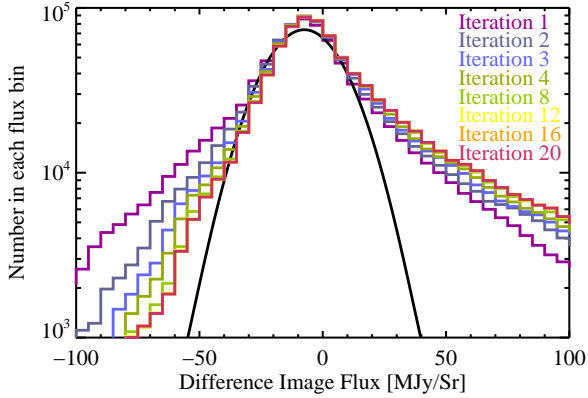}
     \caption{The flux distribution in the difference images (original 500
     \micron~image - convolved cirrus image, e.g. Figure
     \ref{fig:bgtest} panel (f)) for 8 iterations from 1 to
     20 in the $\ell$=30\deg field.  In each iteration, the negative flux 
     distribution was mirrored
     about 0 and a Gaussian was fit to the distribution to determine the
     characteristic fluctuations so that a source identification cutoff
     (of 4.25 $\sigma$) could be applied. Plotted here are the full
     distributions, not the mirrored distributions used to fit the
     Gaussian.  As we iterate, points in the negative end  are transferred 
     to the positive end of the distribution as flux is restored 
     in the negative bowls around bright sources.  The Gaussian fit to the
     distribution for the final iteration (16) is shown as the black
     curve.  While the final distribution is not perfectly Gaussian, the
     iterations show great improvement, and a Gaussian that is characteristic
     of the random fluctuations in the distribution.  We note that the 
     distribution does not peak at zero because the average 
     must be zero and there is a large positive wing in this distribution.}
     \label{fig:itercutoff}
   \end{figure}

We apply a cutoff to this first guess cirrus-subtracted source map, such that
everything above the cutoff is considered `source.'  We determine this
cutoff by fitting a Gaussian to the histogram of pixel values (as shown in
Figure \ref{fig:itercutoff}).  The negative flux values in the distribution
of this ``difference image'' are representative of the random fluctuations
in the data, so we mirror the negative flux values about zero, fit a
Gaussian to that distribution (representative of the noise in the map) and
then apply a cutoff of 4.25 $\sigma$ to the ``difference image.''
We calculated this cutoff in quarter
$\sigma$ intervals from 3 to 6 $\sigma$, a range over which it grows
smoothly.  The choice of 4.25 $\sigma$ was selected because that
cutoff best represented the sources in both the $\ell$=30\deg and $\ell$=59\deg
fields when inspected by eye.  While the choice of cutoff is important
in determining the final physical properties, there is nothing special
about this cutoff; the properties vary smoothly above and below this
value.  

\begin{figure}
     \centering
     \includegraphics[width=0.46\textwidth]{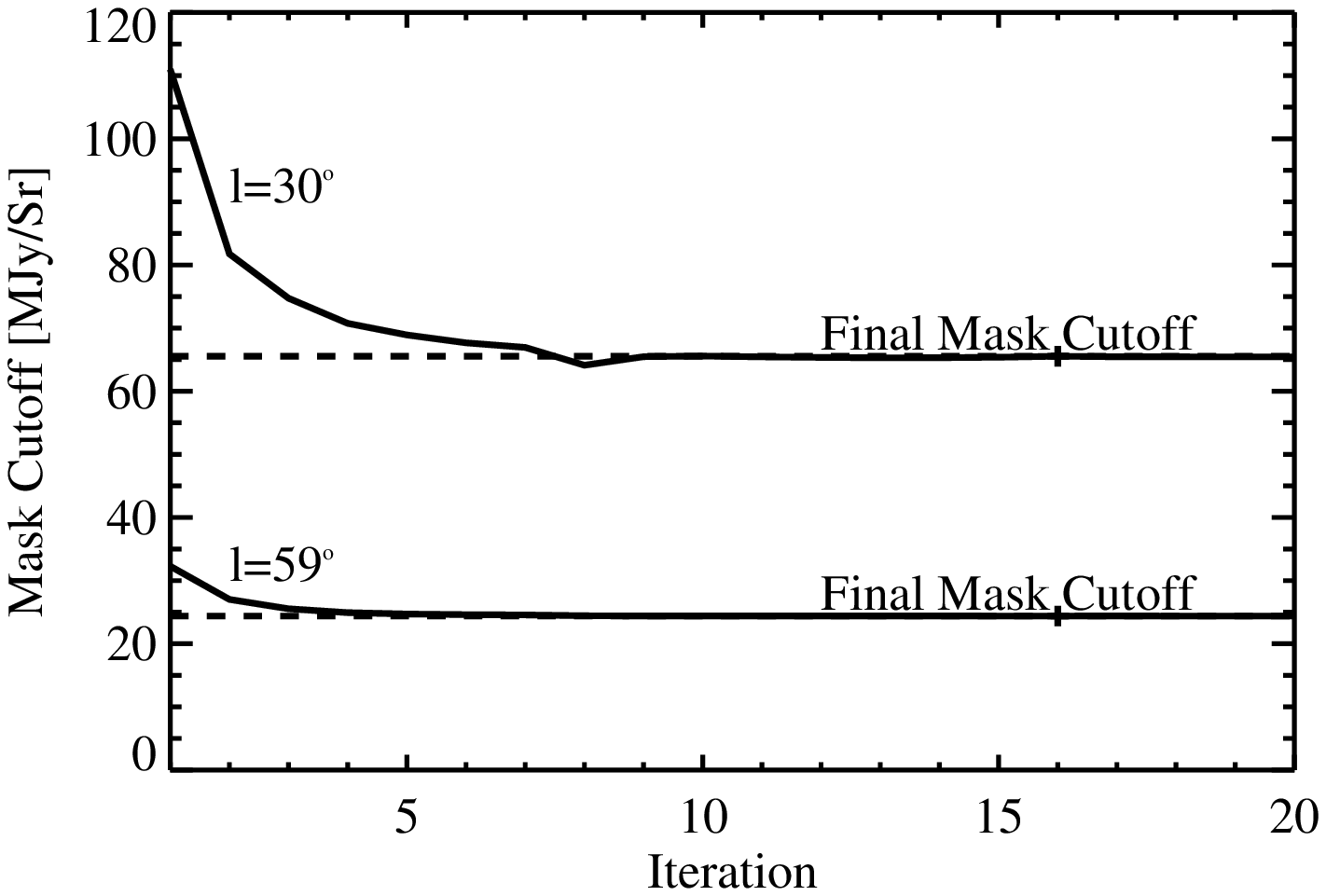}
     \caption{The 4.25 $\sigma$ mask cutoff over 20 iterations.  The cutoff
     is from the Gaussian fit (see Figure \ref{fig:itercutoff}) to the flux
     distribution of points in the difference image (e.g Figure
     \ref{fig:bgtest} panel (f)).  Both the $\ell$=30\deg and $\ell$=59\deg fields
     converge after several iterations, but to very different values.  We
     chose iteration 16 as the final cutoff as it is high enough to be
     assured of convergence and is representative of the value to which the
     iterations converge.}
     \label{fig:iterconv}
   \end{figure}

Once the source cutoff has been determined, the sources are masked out
in the original 500 \micron~image (as shown in solid white in panel (d) of
Figure \ref{fig:bgtest}) and we convolve (with a Gaussian of FWHM 12')
the data outside the source masks, the cirrus emission, treating 
data in the source masks as
missing.  This creates a second guess at the cirrus emission, equivalent to
panel (b).  We then repeat the process of fitting a Gaussian in latitude to
this image (as in panel (c)), determining a source cutoff by fitting a
Gaussian to the mirrored negative flux distribution in the ``difference
image,'' applying these masks (panel (d)), and convolving around the masks
(panel (e)) to produce the next guess at the cirrus emission (panel (b)).
We iterate on this process until the source masks converge.  Figure
\ref{fig:iterconv} shows the convergence of the source cutoff in both
fields.  We chose iteration 16 as the final source mask cutoff as it was
representative of the converged value around which the iterations varied, 
though the
exact choice does not matter much, as the cutoff varies very little after
iteration 10 in both fields.  In panel (d) of Figure \ref{fig:bgtest} the
white masks are the first iteration source masks and the white contour
shows the final (16th) iteration source masks.
Once the source masks are determined, they are applied to each wavelength
image, and that image is convolved (with a Gaussian FWHM 12' and the mask
pixels ignored) to create the cirrus image at that wavelength.  Subtracting
the cirrus image at each wavelength from the original data produces the
source images used in the modified blackbody fits, while the smooth cirrus
images are used for the modified blackbody fits to the cirrus emission. 

The top panel of figure \ref{fig:bgcut} shows a slice of 500 \micron~data 
through Galactic latitude.  The final convolved background (dashed black
line) is representative of the low-lying emission across Galactic latitude,
including it's asymmetry about b=0\deg.  Notice in the bottom panel that
the structure is flat across Galactic latitude and allows for the
identification of the source near b=1\deg.
Figure \ref{fig:itercutoff} shows that
as the iterations converge, we move flux from negative bowls back to
sources as positive features and the mirrored negative distribution then 
more closely resembles a Gaussian (the black line is the final Gaussian 
fit to iteration 16).

This method has been fine-tuned by eye to reproduce the significant
structure picked out by a human, but is entirely automated and was run 
in the exact same way for both the $\ell$=30\deg and $\ell$=59\deg fields, despite their 
vast differences.  We have found that this method faithfully identifies
sources across the range of Galactic latitudes covered by the data and the
range of source confusion and activity observed.  
The sensitivity of this method is dependent upon
the final mask cutoff or confusion, rather than the sensitivity of
Hi-GAL.  For the $\ell$=30\deg final mask cutoff of 65 MJy/sr, we
are sensitive above the cirrus background to a column density of N(H$_{2}$)
= 2.9 $\times$ 10$^{21}$ cm$^{-2}$ for 20 K dust or N(H$_{2}$)
= 9.7 $\times$ 10$^{20}$ cm$^{-2}$ for 40 K dust.  In the $\ell$=59\deg field
the final mask cutoff is about 24 MJy/sr, which gives a sensitivity above
the cirrus background to a column density of N(H$_{2}$)
= 1.1 $\times$ 10$^{21}$ cm$^{-2}$ for 20 K dust or N(H$_{2}$)
= 3.6 $\times$ 10$^{20}$ cm$^{-2}$ for 40 K dust.

\subsection{Modified Blackbody Fits}
\label{sec:graybody}
We performed pixel-by-pixel modified blackbody fits to both the cirrus cloud and the
dense source clump emission.  The dense source pixel-by-pixel fits were first performed on
the images convolved to 36'' resolution.  Following that, the results from
the 36'' resolution tests were used as input to pixel-by-pixel fits
performed on the images convolved to 25'' resolution.  While the fits to
the 25'' resolution images formally utilize less data points, they allow
for a more detailed comparison.

We use the modified blackbody expression in the form
\begin{equation}
\label{eq:graybody}
S_{\nu} = \frac{2h\nu^{3}}{c^{2} (e^{\frac{h\nu}{kT}} - 1)}
(1 - e^{-\tau_{\nu}})
\end{equation}
where 
\begin{equation}
\label{eq:tau}
\tau_{\nu} = \rm \mu_{H_{2}}  m_{H} \kappa_{\nu} \rm N(H_{2}) 
\end{equation}
where $\mu_{H_{2}}$ is the mean molecular weight for which we adopt a 
value of $\mu_{H_{2}}$ = 2.8 \citep{kau08}, m$_{H}$ is the mass of
hydrogen, N(H$_{2}$) is the column density, and $\kappa_{\nu}$ is the dust
opacity.  We determine the dust opacity as a continuous function of
frequency by fitting a power-law of the form \begin{math} \kappa_{\nu} =
  \kappa_{0} (\frac{\nu}{\nu_{0}}) ^{\beta} \end{math} to the tabulated
\citet{oss94} dust opacities in the relevant range of frequencies.  We used
the \citet{oss94} MRN distribution with thin ice mantles that have
coagulated at 10$^{6}$ cm$^{-3}$ for 10$^{5}$ years model for dust 
opacity, which is a reasonable guess for these cold, dense clumps.  We have
assumed a gas to dust ratio of 100, which yielded a
$\kappa_{0}$ of 4.0 at $\nu_{0}$ of 505 GHz with a $\beta$ = 1.75. 

We first fit a modified blackbody to each pixel in the cirrus image (e.g. Figure
\ref{fig:bgtest} panel (e)) using MPFITFUN \citep[from the Markwardt IDL
  Libarary,][]{mar09}.  We assign a calibration uncertainty of 20\% to the
SPIRE and PACS data points and use the covariance matrix returned by
MPFITFUN to estimate errors in the parameters.  In the modified blackbody
fit, we leave the column density, $\beta$, and the
temperature as free parameters.  The temperature was restricted to range
between 0 and 100 K, the column density was free to range between 0 and
100 $\times$ 10$^{22}$ cm$^{-2}$, and $\beta$ from 1 to 3.  We
do not use the BGPS 1.1 mm point in the fits because uncertainties in both
the absolute calibration and the spatial filter function prevent the BGPS
point from significantly helping to constrain the fit at this time.  

We then fit a modified blackbody at each position within the source masks
of the cirrus-subtracted science images (e.g. Figure \ref{fig:bgtest} panel
(f)) to the dense clumps.  For these fits of dense clumps, we ignore the
70 \micron~point since the optically thin assumption may not be valid
($\tau_{70 \mu m}$ = 1 at N(H$_{2}$) = 1.2 $\times$ 10$^{23}$ cm$^{-2}$
using Equation \ref{eq:tau}).  In fact, IRDCs often
appear in absorption at 70 \micron~indicating that in these very high
column density regimes, the 70 \micron~data is not optically thin and 
that it can no longer be modelled by a single dust temperature.  
Additionally, it has been found \citep[e.g.][]{des90, com10} that a large fraction
of the flux at 70 \micron~is likely contaminated by emission from very
small grains whose temperature fluctuations with time are not
representative of the equilibrium temperature of the large grains.
Without the 70 \micron~point we have only four points (usually just 
on the Rayleigh-Jeans slope) to constrain the fit, so we fix
$\beta$ to 1.75 \citep[The][fit value for an MRN distribution with thin ice
mantles that have coagulated at 10$^{6}$ cm$^{-3}$ for 10$^{5}$ years; 
a reasonable guess for these
environments]{oss94}, and leave only the temperature and the column
density as free parameters.  We discuss how a value of $\beta$ of 1.5 or 2
would alter the results in Section \ref{sec:caveats}.  
We acknowledge that perceived changes in temperature may actually be 
changes in $\beta$ and discuss this more also in Section \ref{sec:caveats}.  
We are, to an extent, looking for changes in the dust properties
($\beta$ or temperature) across different environments, which we can still
achieve.  

\begin{figure*}
   \centering
   \includegraphics[width=1\textwidth]{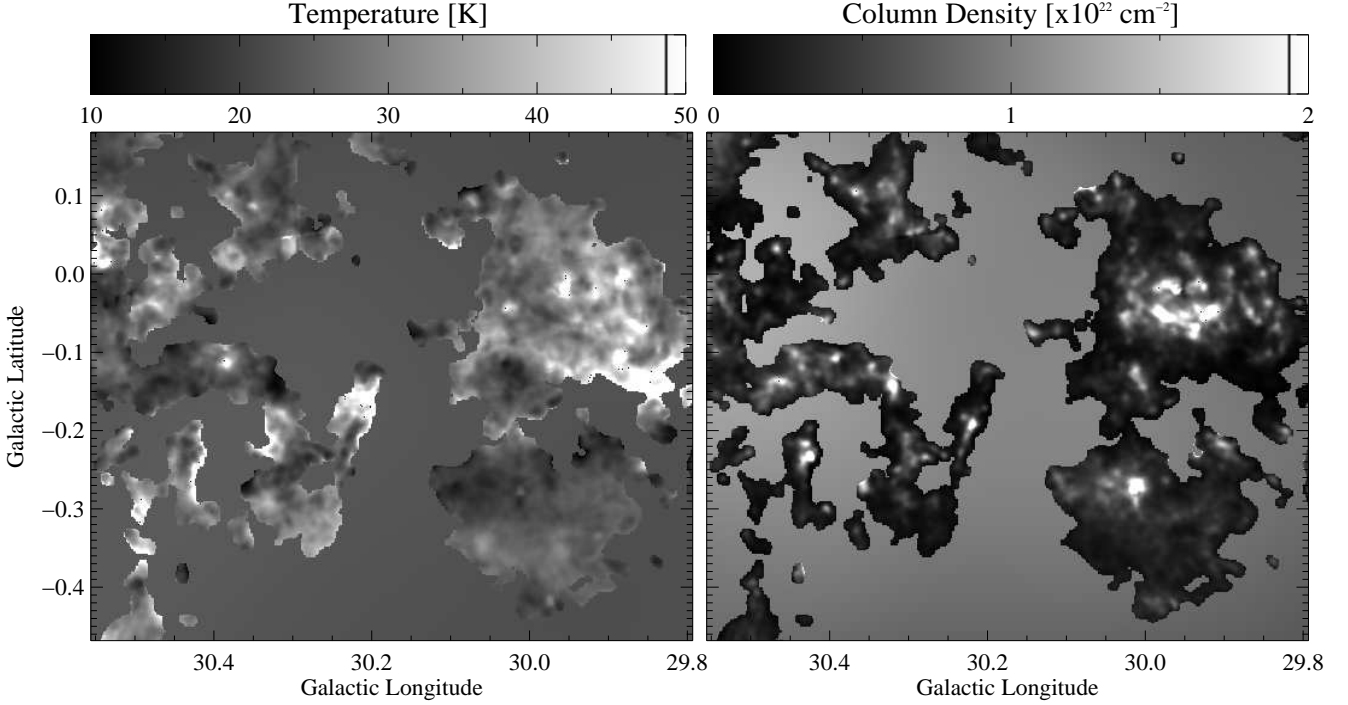}
   \caption{Temperature (left) and Column Density (right) maps in the
     $\ell$=30\deg field test region.  In these maps, the background cirrus
     emission temperature and column densities are plotted outside of the
     source masks, while the background-subtracted temperature and column
     densities are plotted inside the source masks.  These maps are at 25''
     resolution and assume $\beta$ = 1.75 inside the source masks.}
   \label{fig:testirdc_tempcol}
    \end{figure*}

These pixel-by-pixel modified blackbody fits to the cirrus-subtracted dense
clumps were first performed on
the images convolved to 36'' resolution.  Following that, the results from
the 36'' resolution tests were used as input to pixel-by-pixel fits
performed on the images convolved to 25'' resolution.  While the fits to
the 25'' resolution images formally utilize less data points (the SPIRE 500
\micron~point must be excluded because of resolution), they allow
for a more detailed comparison.  The physical properties
returned by the fits at different resolutions agree well.  The final
temperature and column density maps (as shown in Figure
\ref{fig:testirdc_tempcol}) show the derived temperatures and column
densities from the 25'' map within the source masks, and the
smoothed cirrus-background temperatures and column densities outside the
source masks.

We note here that star-forming regions contain structure on many scales, 
and that fitting a single
temperature and column density over a region of 25'' (about 0.5 pc at a
distance of 4 kpc) is a vast oversimplification.  There certainly exist
large variations in the temperature and column density in these sources on
smaller scales.  In this paper, we present the beam-diluted average source
properties on 25'' scales.  The column densities presented here
are therefore characteristic of the larger-scale clump structure and
under-estimates of the peak column densities, while the temperatures will
be underestimates in very hot regions (hot cores) and overestimates in
very cold regions (starless IRDC cores).

\subsection{Star Formation Tracer Label Maps}
\label{sec:sftracers}
In order to robustly compare the physical properties determined in this
paper with various star formation tracers, we have created star formation
tracer label maps.  These maps are on the same grid as the science images
(temperature, column density, etc.) and have a binary denotion in each
pixel: 1 if the star formation tracer is present and 0 if it is not.  We
create star formation tracer label maps for the MIPS 24 \micron~emission, 8
\micron~emission, Extended Green Objects \citep[EGOs;][]{cyg08, cha09}, and
6.7 GHz methanol maser emission \citep{pes05}.  The MIPS 24 \micron~images
required stitching together using MONTAGE before the
label map could be created.  Since the $\ell$=30\deg field is more populated,
confused, and therefore has a wider range of star formation activity, we
only perform the analysis in this field.  The results of this comparison
can  be found in Section \ref{sec:sftracerres}.  The 8 \micron~emission label
maps, however, were created and analyzed for both the $\ell$=30\deg and
$\ell$=59\deg fields so that we could understand the relationship between 
mid-IR and Hi-GAL sources in both fields (see Section \ref{sec:farirmir}).

   \begin{figure}
   \centering
   \includegraphics[width=0.46\textwidth]{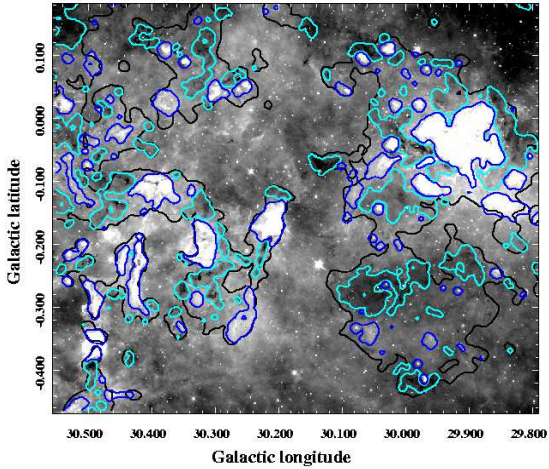}
   \caption{GLIMPSE 8 \micron~image overlaid with contours of the
   classification as mid-IR-bright (mIRb, dark blue) or mid-IR-dark (mIRd, cyan).  The
   classification is only applied where a far-IR source exists, as denoted by
   the black contours.}
   \label{fig:testirdc_glmdiff}
    \end{figure}

\begin{figure}
  \centering
  \includegraphics[width=0.46\textwidth]{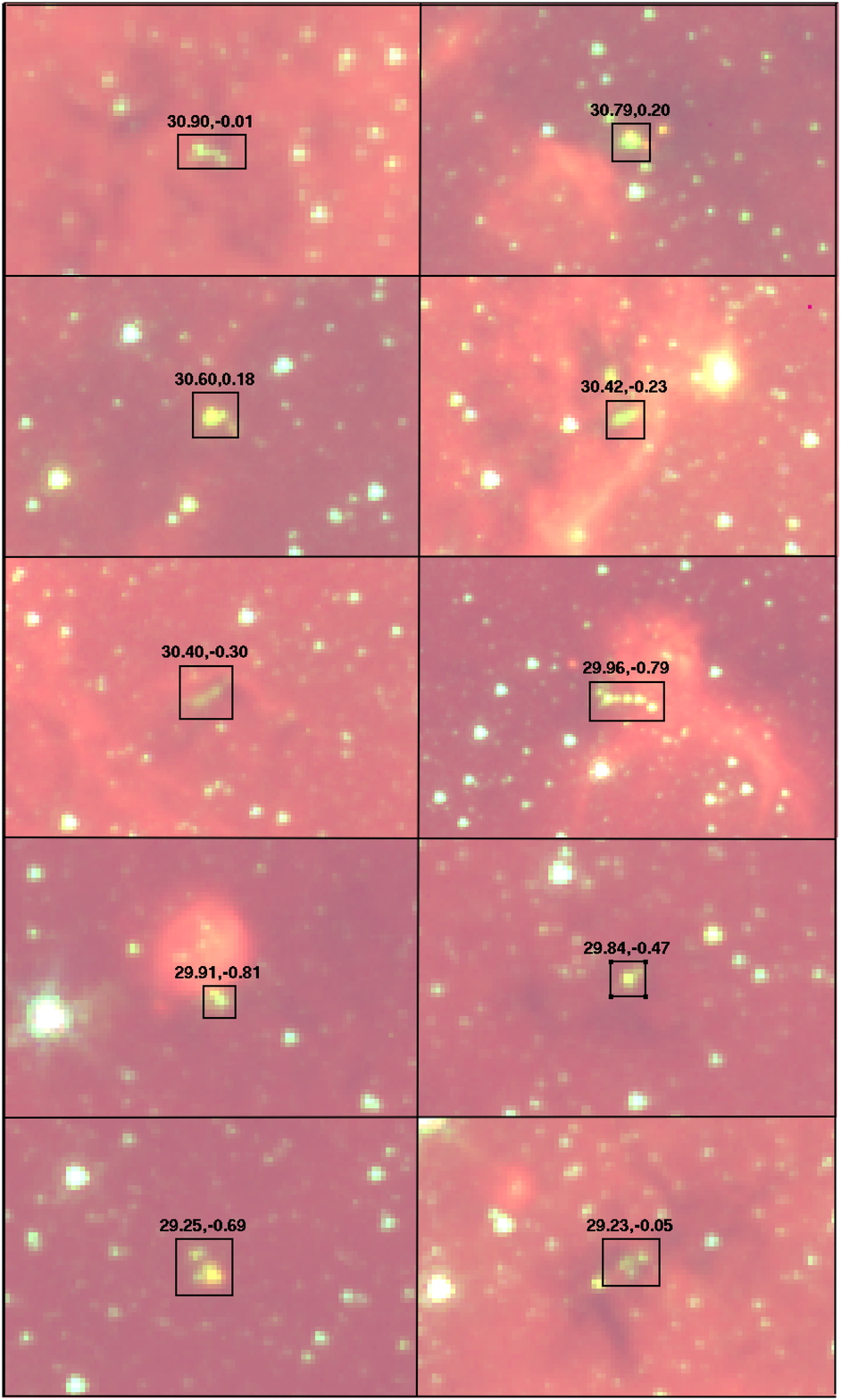}
  \caption{GLIMPSE 3-color images (Red: 8 \micron, Green: 4.5 \micron,
  Blue: 3.6 \micron) of EGOs in the $\ell$=30\deg field.  The EGOs shown
  in the right column of the 3rd and 4th row were identified as
  ``possible'' EGOs by \citet{cyg08} and the others were identified by two
  independent viewers as EGOs as described in Section \ref{sec:sftracers}.
  The positions are listed in Table \ref{tab:egos}.}
  \label{fig:egos}
  \end{figure}

The GLIMPSE 8 \micron~band shows emission from warm dust and Polycyclic
Aromatic Hydrocarbons (PAHs).  Regions with bright 8 \micron~emission 
may contain warm, diffuse dust or UV-excited PAH emission from an
Ultra-Compact HII Region \citep{bat10}.  The absorption of 8 \micron~emission by an
Infrared Dark Cloud (IRDC) is indicative of a high column of cold dust
obscuring the bright mid-IR background.
We note that mid-IR-bright regions are generally indicative of active star
formation, and that the peak of mid-IR light can be offset from the
youngest regions of star formation \citep[e.g.][]{beu07a}.
We denote regions of bright 8 \micron~emission as mid-IR-bright (mIRb),
dark regions of absorption at 8 \micron~as mid-IR-dark (mIRd), and regions
without stark emission or absorption at 8 \micron~as mid-IR-neutral (mIRn).
The 8 \micron~emission label map has both a negative (IRDC, mIRd),
positive (mIRb), and neutral (mIRn) component.  This map is created by
first making a contrast image.  The GLIMPSE 8 \micron~image is convolved
with a Gaussian of FWHM 5' to represent the smooth, slowly varying
background.  5' was determined \citep[see][for more details on this smoothing kernel
choice]{per09} to be large enough to capture the largest
IRDCs and mid-IR-bright objects, while still small enough to capture the
background variations.  This image is then subtracted from the original image, which
is then divided by the smoothed background.  The resulting `contrast map' enhances the
contrast of mIRb and mIRd clouds above the bright mid-IR background.  To obtain 
a fair comparison with the lower resolution ($\sim$ 25'') science images,
this contrast image is convolved to the same resolution.  We then apply a
cutoff such that any pixel with a positive contrast greater than or equal
to 10\%  is considered mIRb, and any pixel with a negative contrast
greater than or equal to 5\% is considered mIRd, while all others are then
mIRn.  We rely on our eyes to pick out regions of emission and absorption,
and have selected the cutoffs that best represent a visual interpretation of
that which is dark and bright, as shown in Figure
\ref{fig:testirdc_glmdiff}. 

A portion of this label map is shown in Figure \ref{fig:testirdc_glmdiff}.
This method does a very good job of picking out the
bright and dark regions, and agrees reasonably well (considering the
different methods, resolution used, and cutoffs) with the IRDC catalog of
\citet{per09}.  Two important biases to note with this method are: 1) the
contrast image technique creates `negative bowls' around bright objects,
and therefore the sizes of the mIRd and mIRb regions are underestimated,
and 2) convolving the contrast image to 25'' resolution causes us to miss
some small IRDCs.  Additionally, any decrements in the background will be
denoted as being mIRd.  This issue is partly resolved when we multiply this 
label map with our source masks, so that our classifications only apply 
where a Hi-GAL source also exists.

Emission at 24 \micron~observed using the MIPS instrument traces warm
dust.  MIPS 24 \micron~emission can trace warm, diffuse dust in the ISM or
warm dust associated with star formation such as material accreting onto forming
stars, the gravitational contraction of a young stellar object,  
or warm dust surrounding a newly formed star.
The MIPS 24 \micron~emission label map was created using the same method as
the 8 \micron~label maps, with the exception that we only include a
positive contrast cutoff.  While some IRDCs remain dark at
24 \micron, the negative contrast is very minimal when convolved to 25''
resolution.  Additionally, the highest negative contrast in the contrast 
image is from the negative bowls around bright sources, so identifying
IRDCs in the 24 \micron~images using this method is not adequate.
The positive contrast threshhold for the MIPS 24 \micron~contrast maps was
25\%, also chosen by eye.  A higher cutoff than at 8 \micron~was
required to select the 24 \micron~sources above the bright background.
These cutoffs greatly affect the resulting trends.  See Section
\ref{sec:caveats} for a discussion of the effects of cutoffs on source
identification. 

We include the presence of Extended Green Objects \citep[EGOs, also called
``green fuzzies''][]{cyg08, cha09} as a star formation tracer.  EGOs are
regions of enhanced and extended 4.5 \micron~emission and are thought to be
indicative of shocks in outflows \citep{cyg09}.  \citet{cyg08} catalogued 
EGOs throughout the Galaxy in the GLIMPSE fields.  There are two
``possible'' EGOs in the $\ell$=30\deg field which we include in our EGO label
map.  We supplemented the \citet{cyg08} catalog with an independent visual
search of the $\ell$=30\deg field for EGOs that may have been missed.  An
independent, cross-checked search identified eight further candidates,
which we include in our analysis.  A list of the locations
of the EGOs used in this analysis is given in Table \ref{tab:egos}, and
they are shown in Figure \ref{fig:egos}.  For each EGO, a box was drawn
around the region that contains the EGO, and each pixel in this box was
given a flag of 1, while for pixels outside the box the flag is 0.  

Finally, we employ the presence or absence of 6.7 GHz Class II CH$_{3}$OH
masers as a star formation tracer.  We use the complete catalog compiled by
\citet{pes05} to identify unbiased searches for 6.7 GHz methanol masers in
the $\ell$=30\deg field.  While there have been numerous targeted searches for
CH$_{3}$OH masers, we utilize the unbiased Galactic Plane searches of
\citet{szy02} and \citet{ell96}, on the 32m Torun Radio Telescope (FWHM of
5.5') and the University of Tasmania 26m Radio Telescope (FWHM of 7'),
respectively.  6.7 GHz methanol masers are found to be exclusively
associated with massive star formation \citep{min03}, and are often offset 
from the radio continuum emission indicative of an
UCHII Region \citep[e.g.][]{wal98,min01} and are thought to represent an
earlier stage of massive star formation.  We chose to include only the
unbiased catalogs of \citet{ell96} and \citet{szy02} so as not to choose
mIRb regions preferentially.  These catalogs together give 22 methanol
maser regions, which we use to create a label map where the ``size'' of the
methanol maser regions is given by $2 \times$ the positional accuracy RMS
of the observations 
\citep[30'' and 36'' for][respectively]{szy02, ell96}.  Methanol maser
emission comes from very small areas on the sky
\citep[e.g.][]{wal98,min01}, however, methanol masers are often clustered
\citep[][find multiple maser spots towards the majority of sources]{szy02,
  ell96thesis}, so these ``source label sizes'' are meant to be
representative of the extent of the star-forming region.

\begin{table}
\caption{EGOs Used in This Paper}
\label{tab:egos}
\centering
\begin{tabular}{lcccc}
\hline\hline
EGO Name & R.A. (J2000) &
Decl. (J2000) & l x b Size('') &
Ref.\tablefootmark{a} \\
\hline
G29.23-0.05  & 18:44:51.4  &  -03:18:41.7  &  14 x 12  &  1  \\
G29.25-0.69  & 18:47:10.7  &  -03:35:42.0  &  14 x 14  &  1  \\
G29.84-0.47  & 18:47:28.8  &  -02:58:03.0  &  9 x 9    &  2  \\
G29.91-0.81  & 18:48:47.6  &  -03:03:31.1  &  8 x 8    &  1  \\
G29.96-0.79  & 18:48:49.8  &  -03:00:28.7  &  32 x 32  &  2  \\
G30.40-0.30  & 18:47:52.6  &  -02:23:12.2  &  17 x 17  &  1  \\
G30.42-0.23  & 18:47:40.8  &  -02:20:31.6  &  12 x 12  &  1  \\
G30.60+0.18  & 18:46:33.7  &  -01:59:30.3  &  14 x 14  &  1  \\
G30.79+0.20  & 18:46:48.2  &  -01:48:54.4  &  12 x 12  &  1  \\
G30.90-0.01  & 18:47:45.9  &  -01:49:00.0  &  14 x 7   &  1  \\
\hline
\end{tabular}
\tablefoot{
  \tablefoottext{a}{1 - This work (see Section \ref{sec:sftracers} and
  Figure \ref{fig:egos}), 2 - \citet{cyg08}} }
\end{table}

   \begin{figure*}
     \centering
     \subfigure{
       \includegraphics[width=0.42\textwidth]{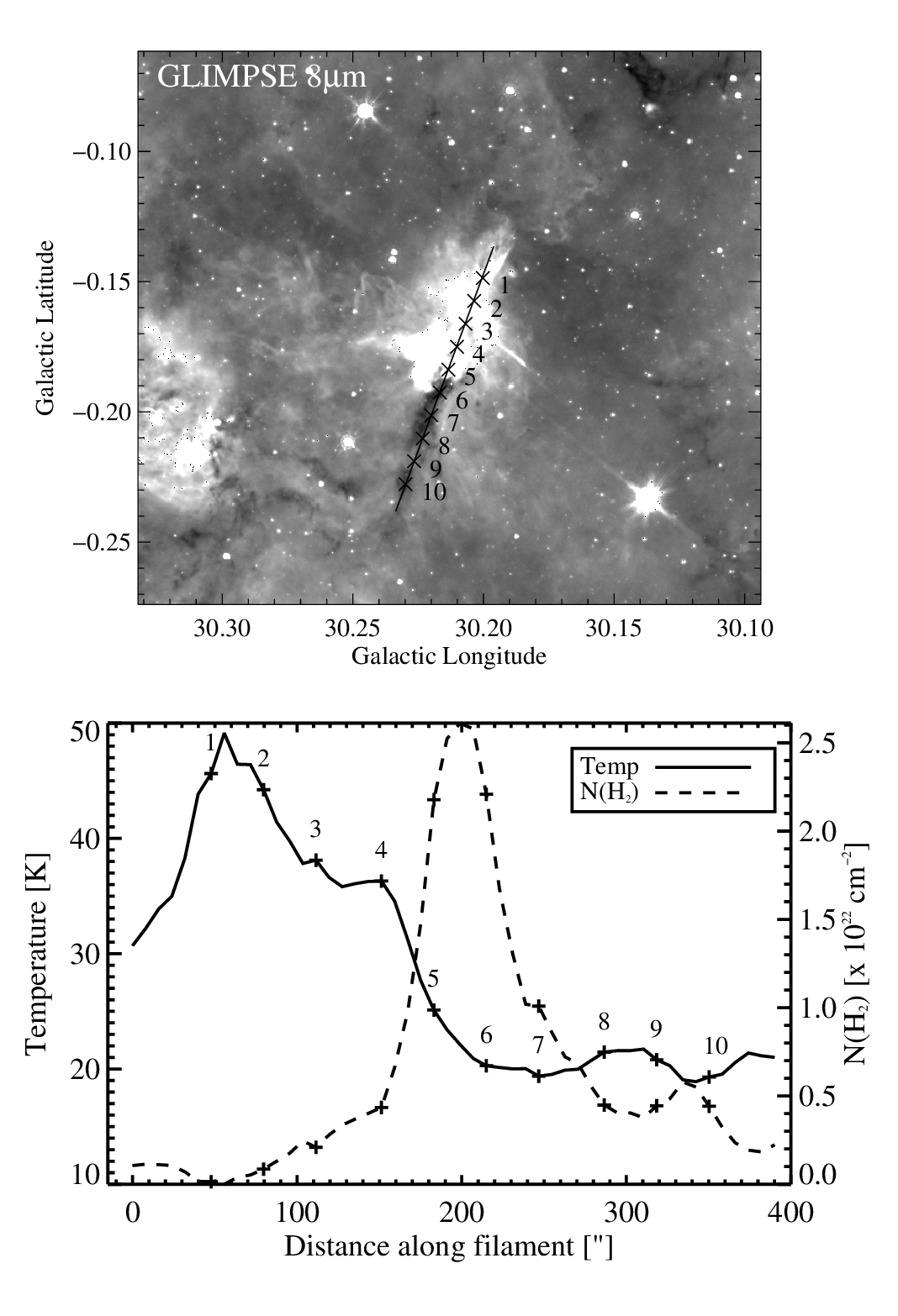}}
     \subfigure{
       \includegraphics[width=0.5\textwidth]{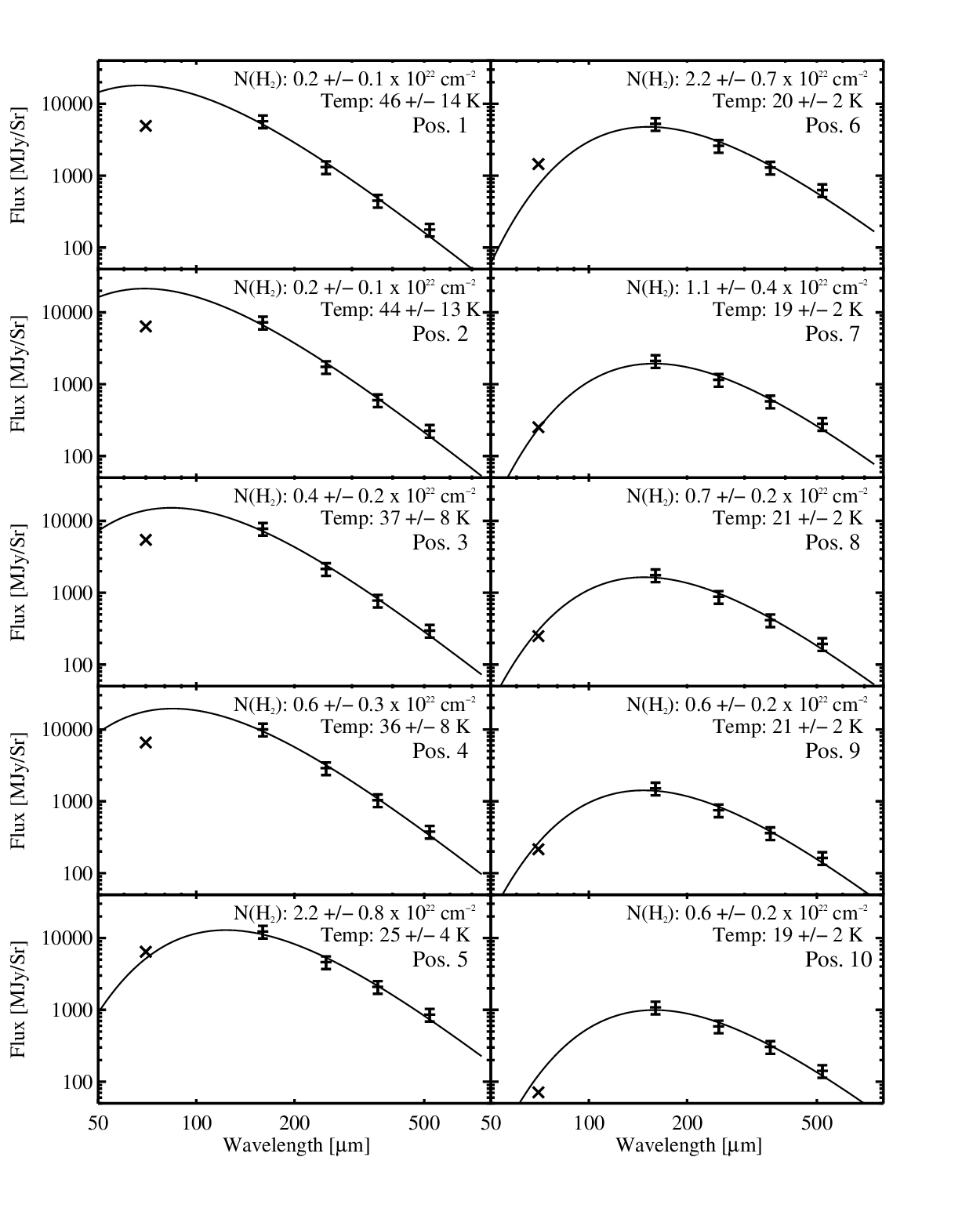}}\\
     \caption{\textit{Top Left:} GLIMPSE 8 \micron~image in our $\ell$=30\deg
     test region.  \textit{Right:} Modified blackbody fits to the points
     marked on the filament in the top left panel.  Since the 70
     \micron~point is not included in these fits (see Section
     \ref{sec:graybody}) it is plotted as an X while the other points are
     shown with their 20\% calibration error bars.  The errors quoted in
     temperature and column density are the formal fit errors.  As
     discussed in Section \ref{sec:caveats}, these are not necessarily
     representative of the true errors.  \textit{Bottom
     Left:} The temperature and column densities derived along the
     filament, with 10 points marked.  Note the inverse relation
     between temperature and column density in this example, especially
     from points 8 to 10.}  
    \label{fig:filament_temp}
     \end{figure*}

   \begin{figure*}
     \centering
     \subfigure{
       \includegraphics
	   [width=0.46\textwidth]{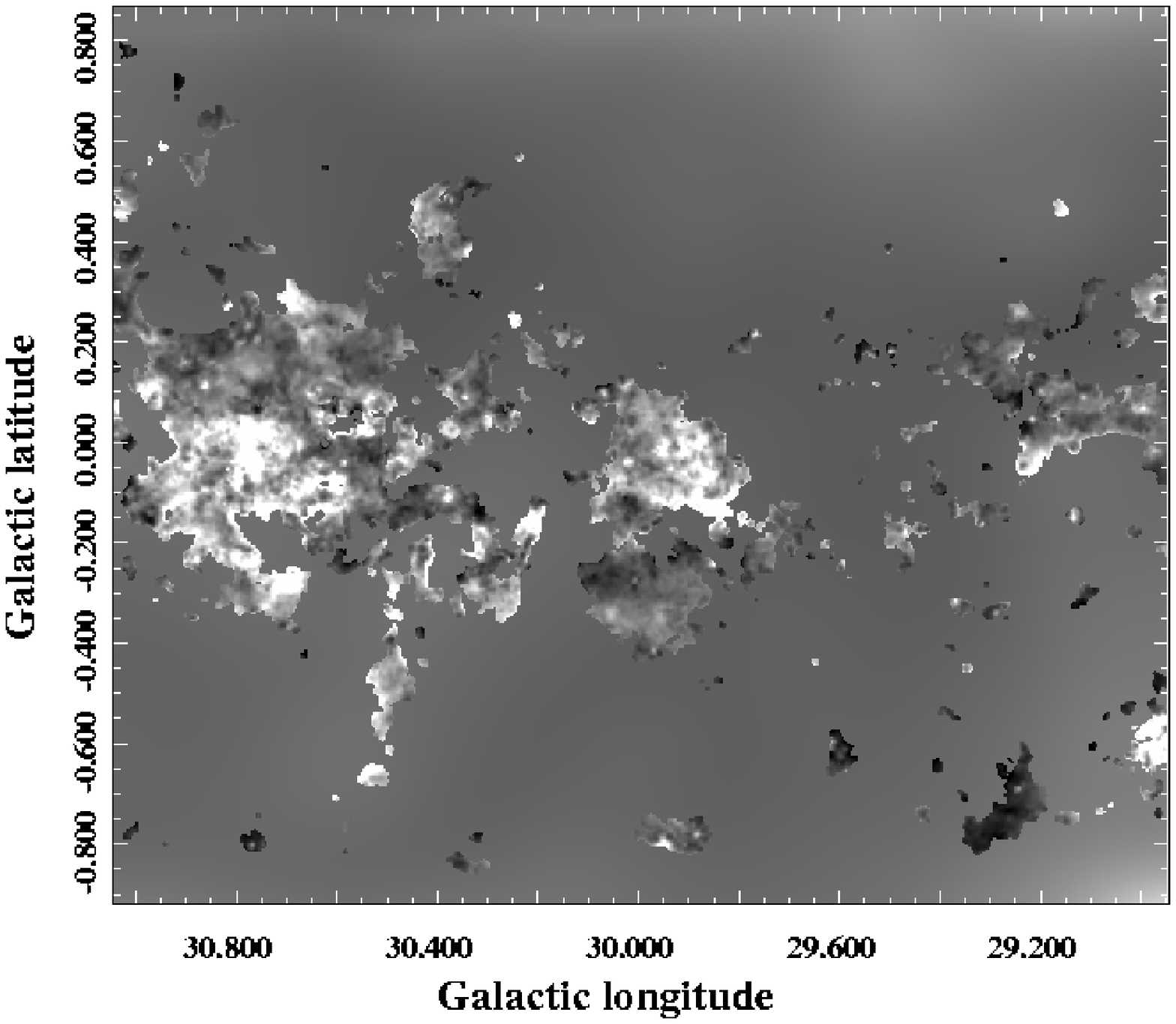}}
    \subfigure{
       \includegraphics
	   [width=0.46\textwidth]{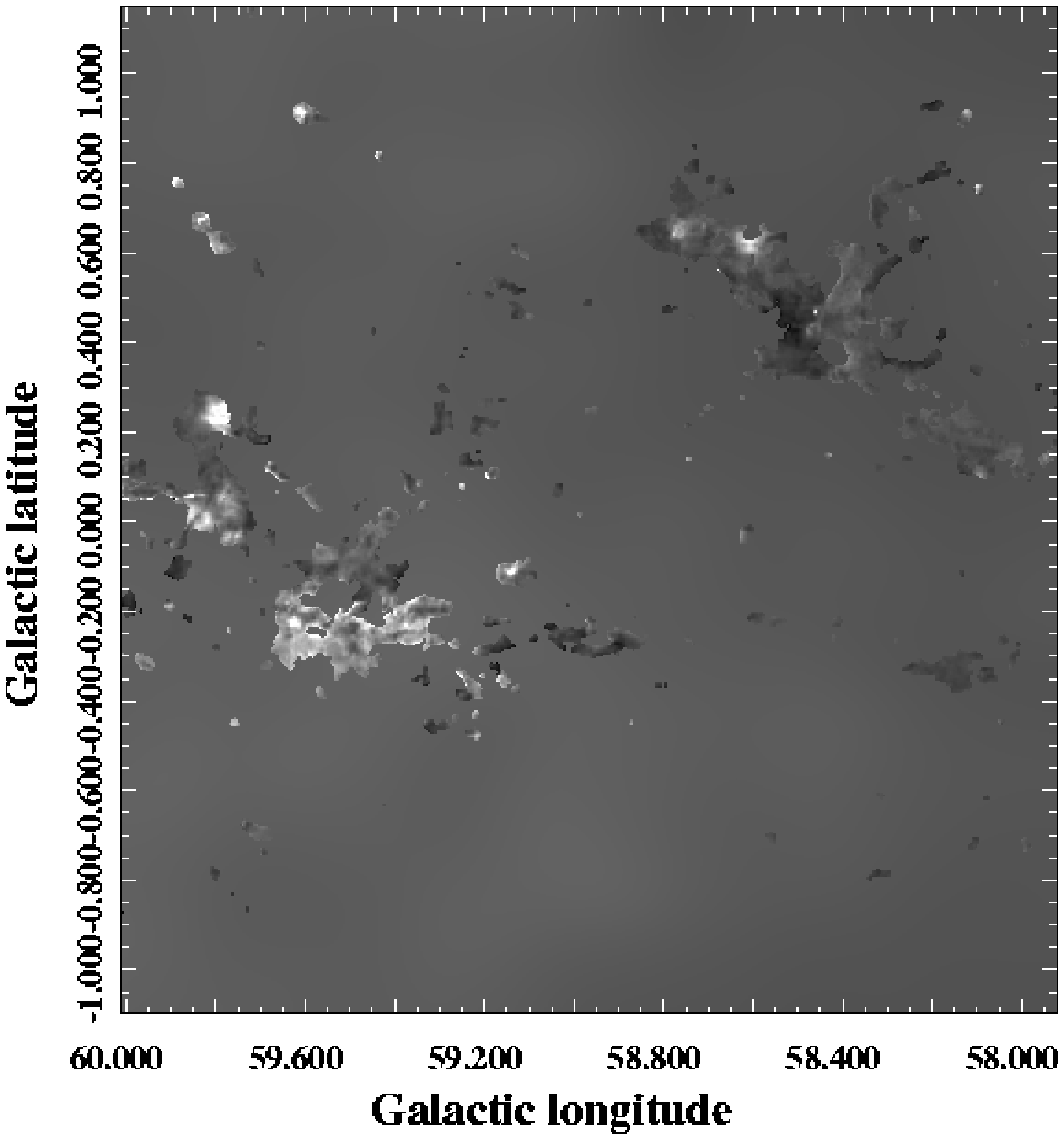}}\\
     \subfigure{
       \includegraphics
	   [width=0.46\textwidth]{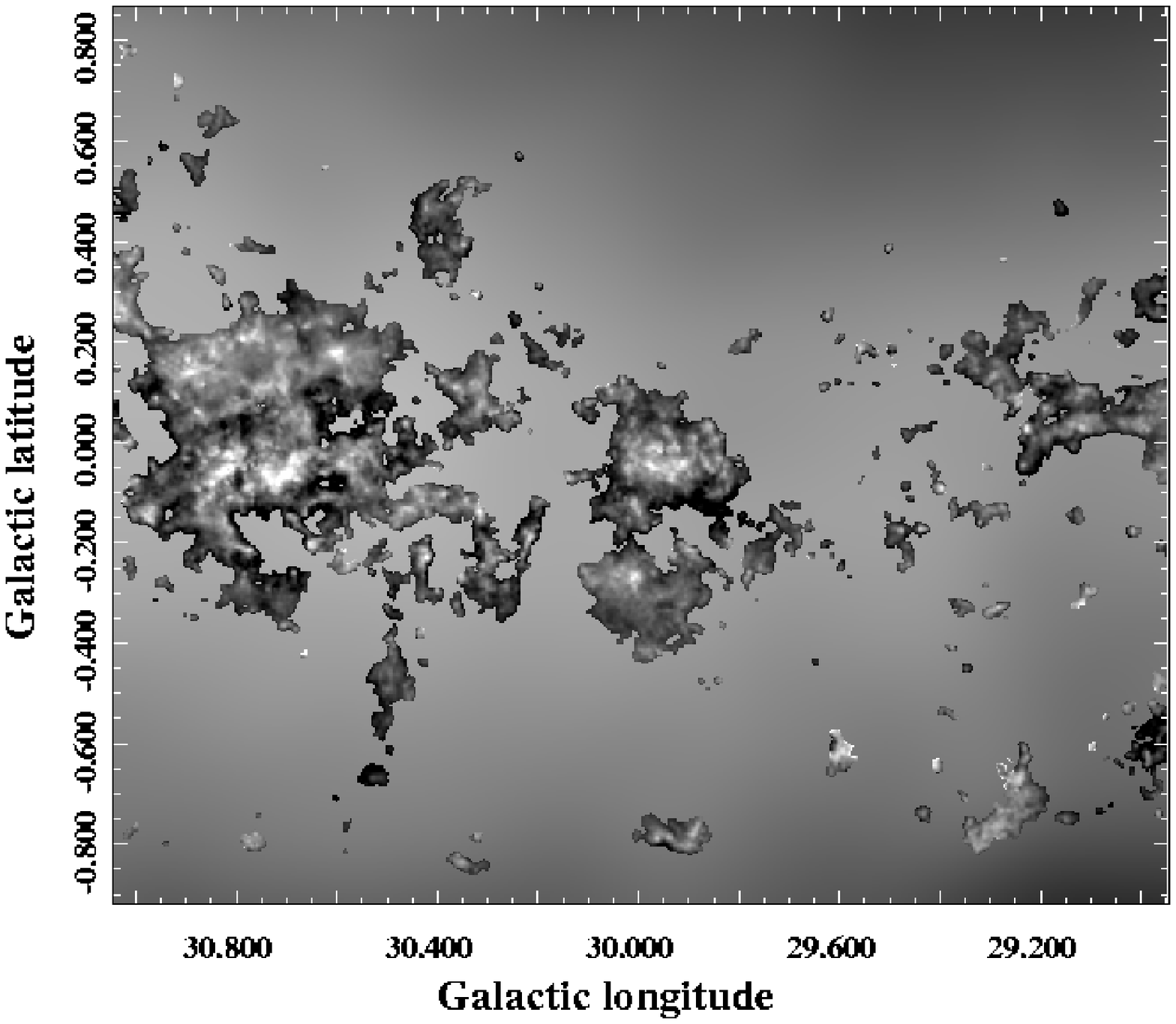}}
     \subfigure{
       \includegraphics
	   [width=0.46\textwidth]{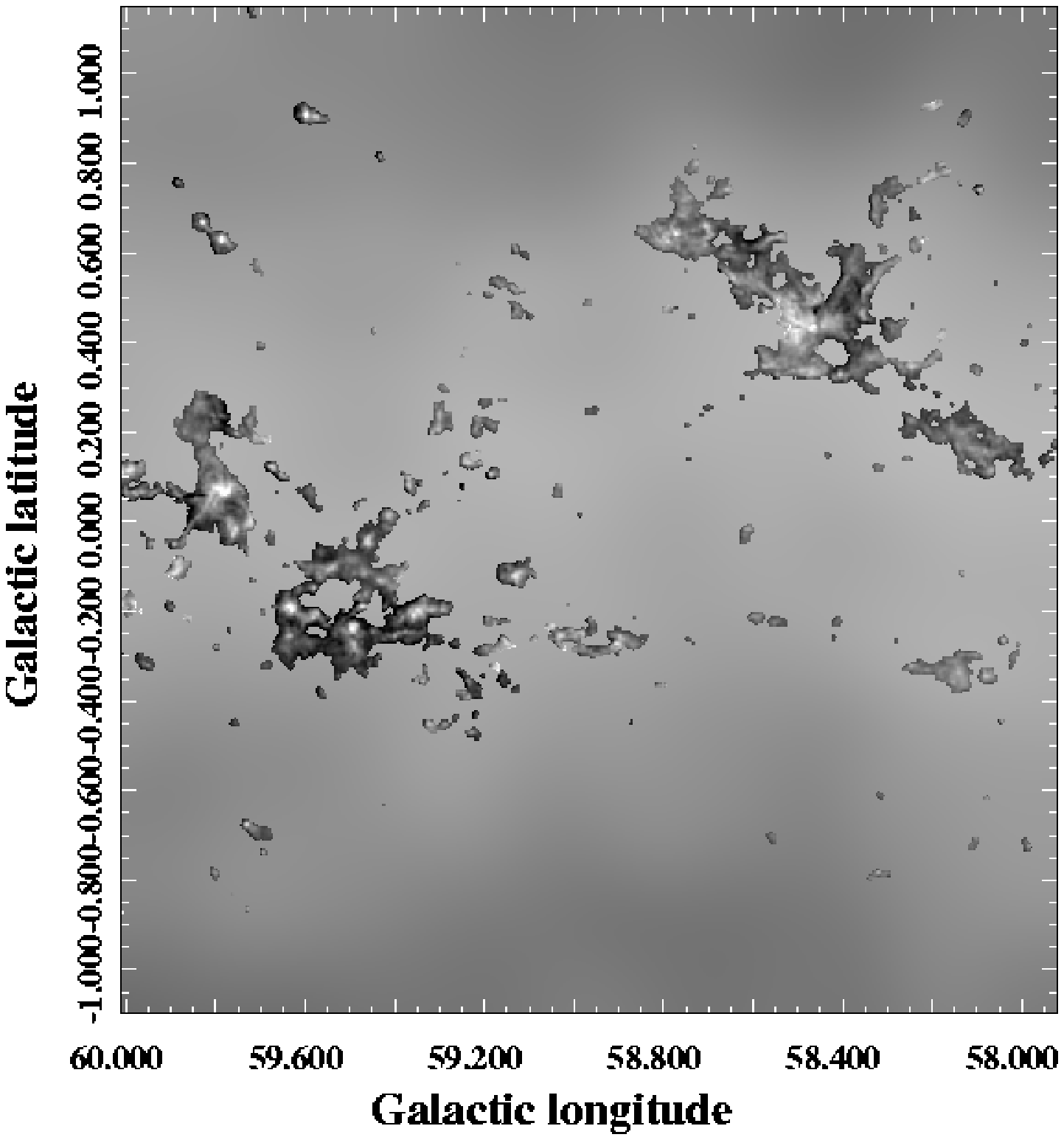}}\\
     \caption{\textit{Top:} Temperature (on a linear scale from 0 to 50 K)
     and \textit{Bottom:} Column Density (on a log scale from N(H$_{2}$)= 0
     to 5 $\times$ 10$^{22}$ cm$^{-2}$) in the $\ell$=30\deg (left) and $\ell$=59\deg
     (right) fields. 
     In these maps, the background cirrus
     emission temperature and column densities are plotted outside of the
     source masks, while the background-subtracted temperature and column
     densities are plotted inside the source masks.  These maps are at 25''
     resolution and assume $\beta$ = 1.75 inside the source masks.}
    \label{fig:all_tempcol}
     \end{figure*}

\begin{figure*}
  \centering
  \subfigure{       
    \includegraphics[trim = 10mm 10mm 5mm 10mm, width=0.49\textwidth]{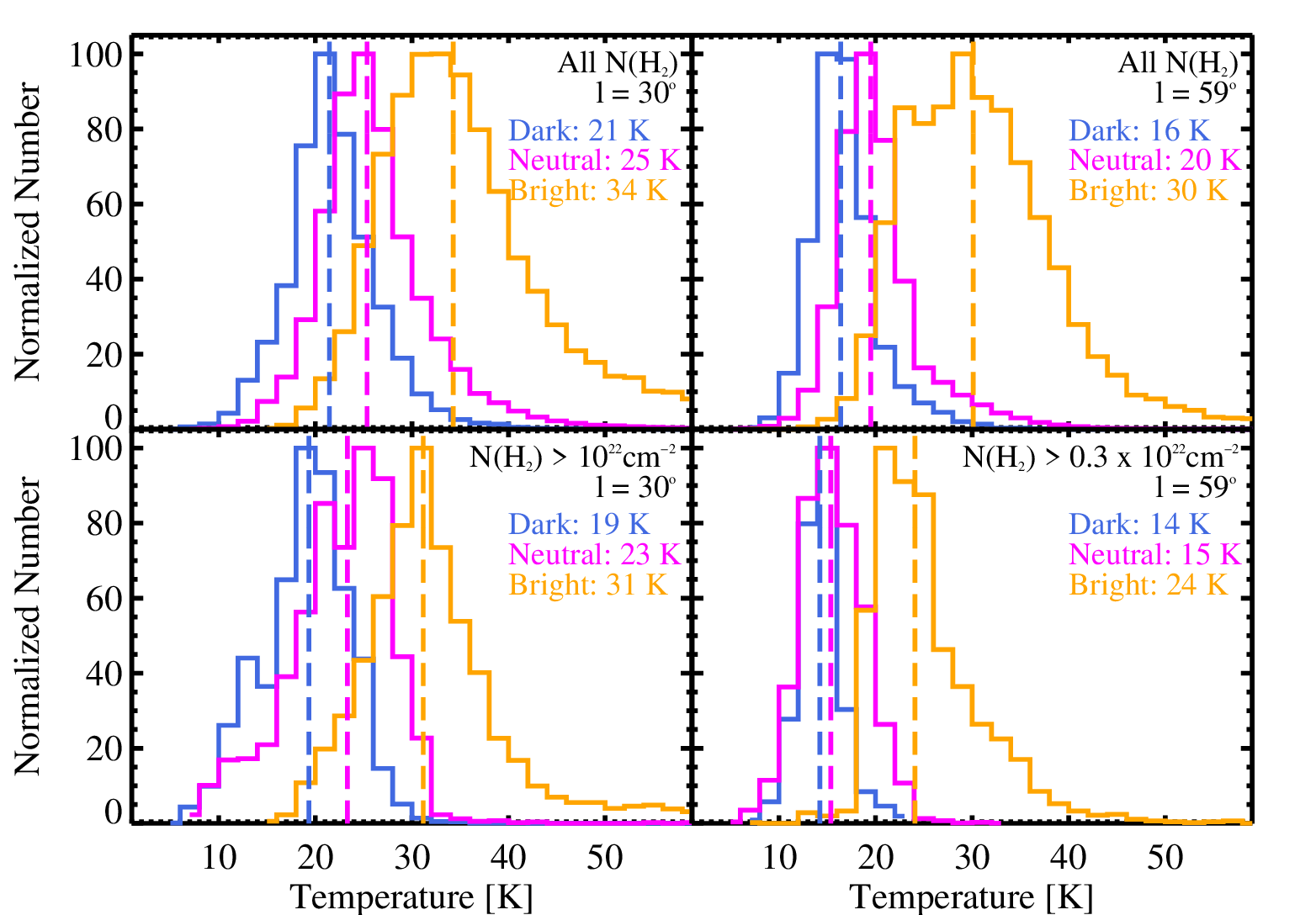}}
  \subfigure{
    \includegraphics[trim = 5mm 10mm 10mm 10mm, width=0.49\textwidth]{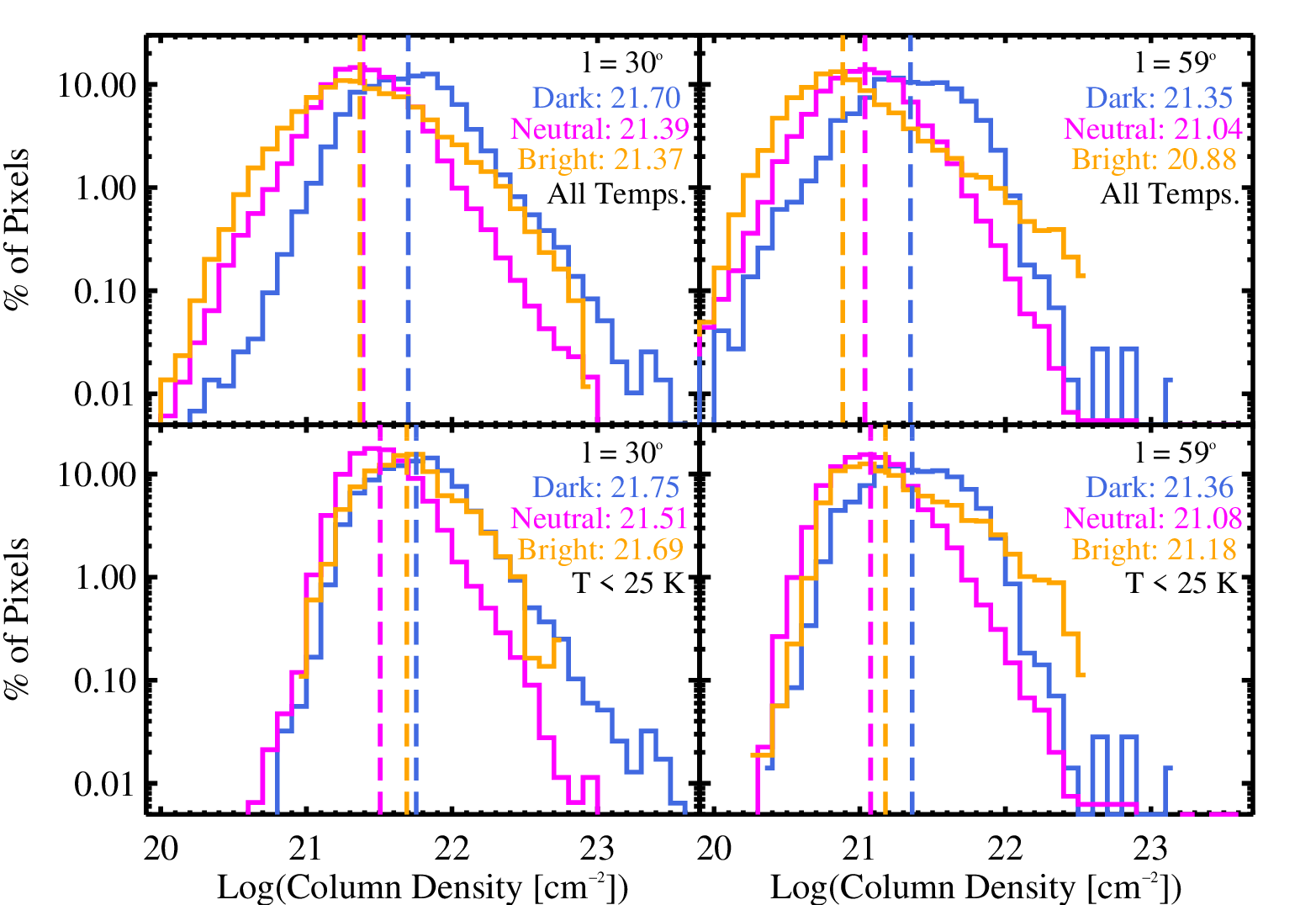}}\\
  \caption{Normalized temperature and column density histograms of source
  pixels in the Hi-GAL fields separated by their 8 \micron~association;
  mid-IR-dark (mIRd), mid-IR-neutral (mIRn), or mid-IR-bright (mIRb).
  The values printed on the plots are the medians of the distributions 
  (also seen in Table \ref{table-midIR}). The mIRd pixels tend
  to have the lowest temperature and highest column density, while the
  mIRb pixels have the highest temperature and lowest column density, and
  the mIRn pixels fall in the middle.  A simple K-S test (see Section
  \ref{sec:farirmir}) shows that the hypothesis that any of these
  populations (in temperature or column density, but not with any cutoffs,
  so the top panels only) are drawn from the same 
  distribution can be ruled out with
  at least 99.7\% confidence.  The average temperature and
  column density is lower in the $\ell$=59\deg field than in the $\ell$=30\deg field.
  \textit{Left:} Normalized temperature
  histograms in the $\ell$=30\deg (left) and $\ell$=59\deg fields (right).  The top
  panels are the temperature histograms of all source pixels in the 25''
  resolution images.  The bottom panels are the temperature histograms of
  all source pixels above a column density cutoff (N(H$_{2}$) $> 10^{22}
  \rm{cm}^{-2}$ in the $\ell$=30\deg field, N(H$_{2}$) $> 0.3 \times 10^{22}
  \rm{cm}^{-2}$ in the $\ell$=59\deg field).  Introducing a column density
  cutoff decreases the average temperature, especially for the mIRb pixels.  
  \textit{Right:} Normalized percentage logarithmic column density
  histograms in the $\ell$=30\deg (left) and $\ell$=59\deg fields (right).  The top
  panels are the column density histograms of all source pixels in the 25''
  resolution images.  The bottom panels are the column density histograms of
  all source pixels with a temperature less than 25 K.  These low
  temperature pixels have higher column densities on average, especially
  the mIRb ones.}
  \label{fig:8micron_hist}
\end{figure*}

  \begin{figure*}
     \centering
     \includegraphics[width=0.96\textwidth]{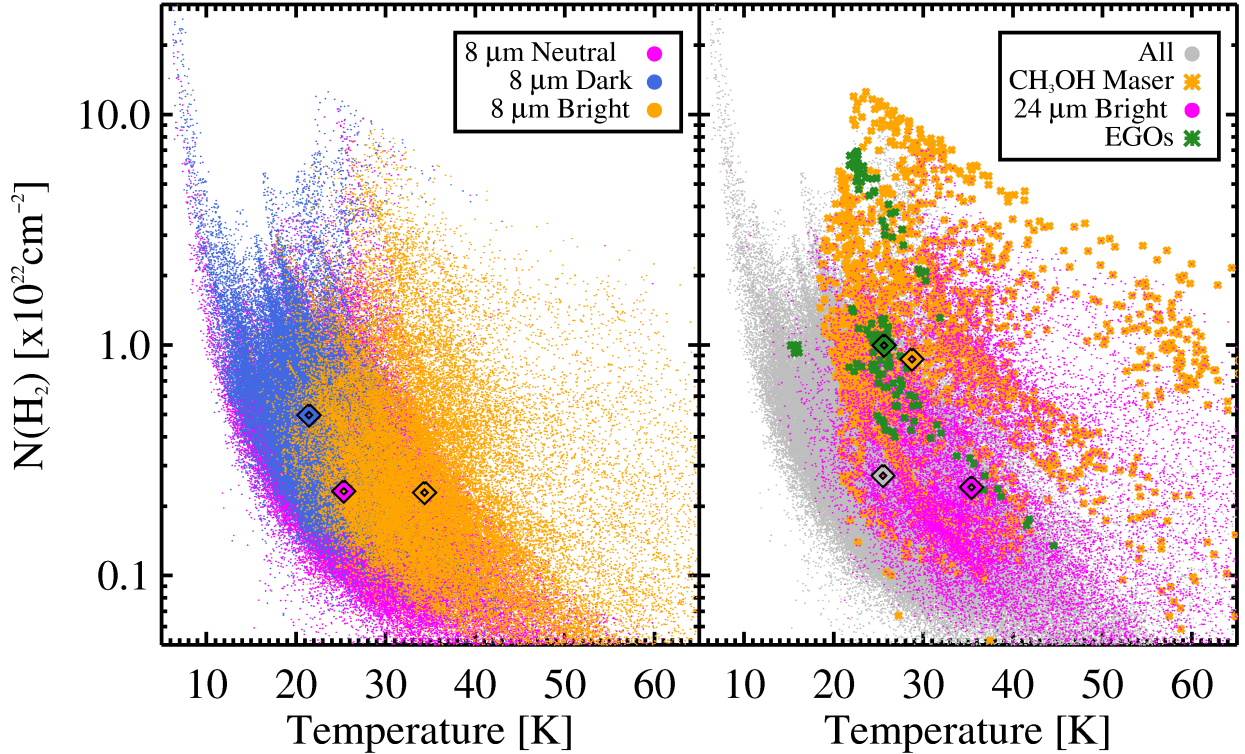}
     \caption{\textit{Left:} A column density versus temperature plot
     comparing mIRn (8 \micron~neutral, magenta), mIRd (8 \micron~dark,
     blue), and mIRb (8 \micron~bright, orange) pixels within the $\ell$=30\deg
     source masks, with the median of each value plotted as a diamond on
     top.  The mIRd points occupy the coldest, highest column density
     portion of the plot, while the mIRb points occupy a range of column
     densities.  The mIRb points average a lower column density and a higher
     temperature region of the plot than the mIRd points. The mIRn points average a lower
     column density than mIRd points, and a slightly higher temperature,
     but otherwise match that distribution somewhat.  The mIRn points at
     the extreme left of the plot are due to the imperfect definition of
     the 8 \micron~masks.  Where mIRb and mIRd regions are adjacent,
     the pixels between are denoted as mIRn since the two adjacent regions
     are blurred.  This causes some few pixels to be denoted as mIRn which may
     truly be mIRd or mIRb.  \textit{Right:} A
     similar plot as the left, but with all points (gray), points
     associated with a CH$_{3}$OH maser (orange) or EGO (green), and 24
     \micron~bright points (magenta).  We note that EGOs occupy a mid-range
     temperature and the highest column density portion of this plot, while
     the CH$_{3}$OH maser points occupy a wide mid-range of temperatures
     and a high range of column densities.  The 24
     \micron~points are associated with higher temperature points and
     relatively low column density.}  
     \label{fig:l30_tempcol_sf}
   \end{figure*}

   \begin{figure}
     \centering
     \subfigure{
       \includegraphics[width=0.46\textwidth]{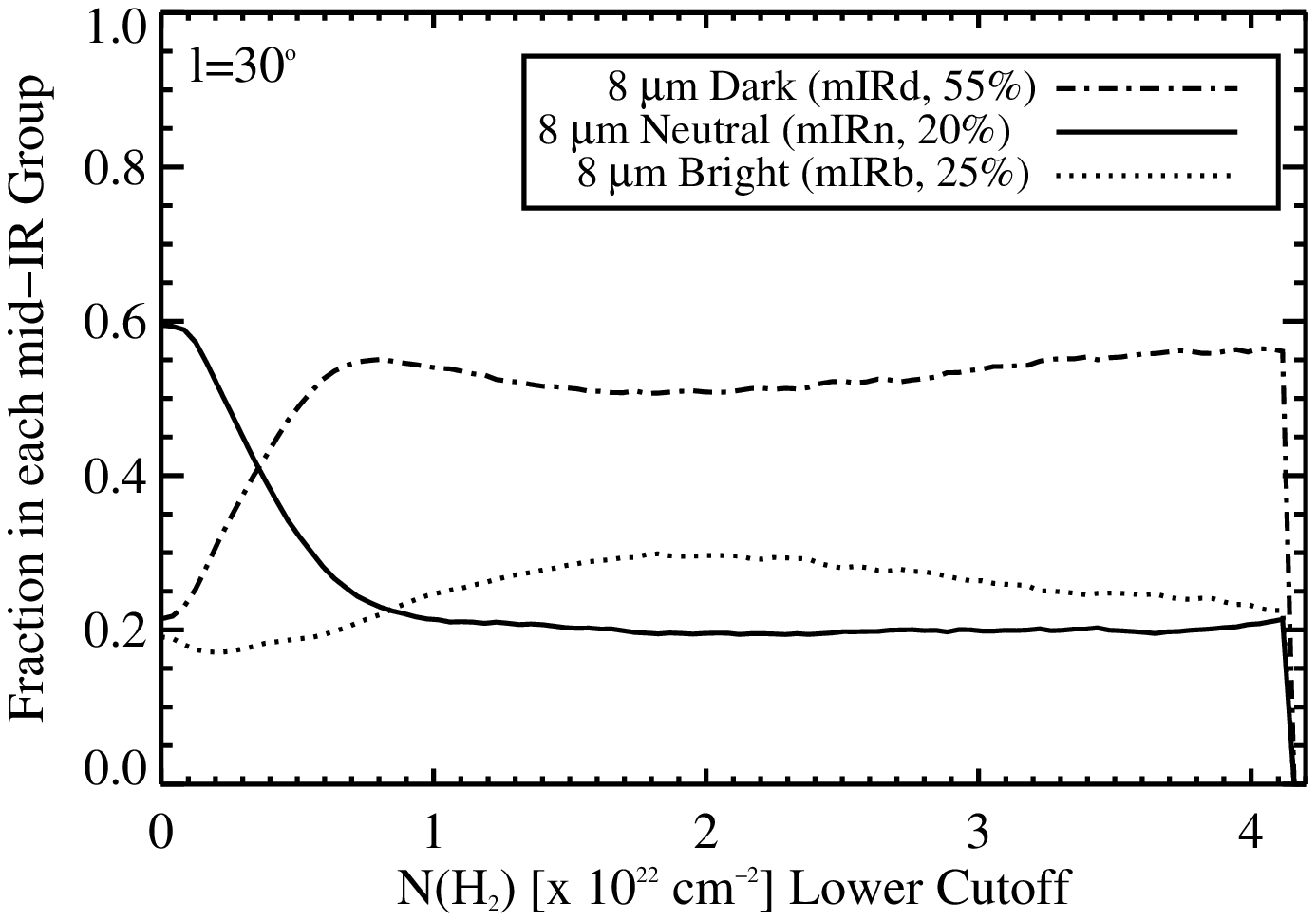}}\\
     \subfigure{
       \includegraphics[width=0.46\textwidth]{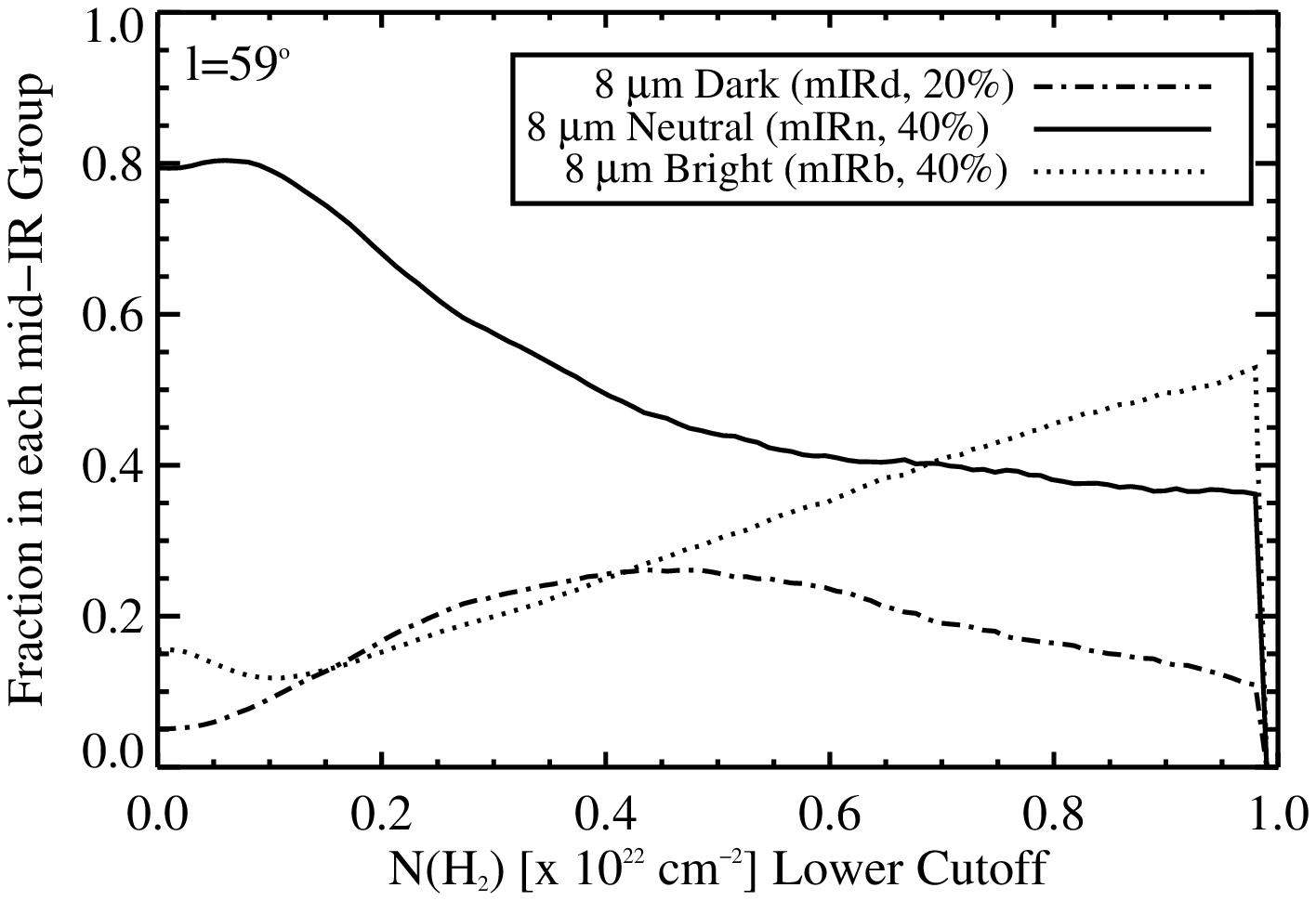}}\\
     \caption{Fraction of pixels in each mid-IR categorization (mIRd, mIRn,
       mIRb) as a function of column density cutoff for the $\ell$=30\deg (top)
       and $\ell$=59\deg fields.  In the $\ell$=30\deg field, as the column density
       cutoff is increased, the mIRd pixels become much more prevalent,
       replacing the common, but low-significance mIRn pixels.  In the
       $\ell$=59\deg field, the mIRn contribution again lowers, while in this
       case, the mIRb pixels become more prevalent.  In both plots, the
       sharp cutoff near 4 and 1 indicated where the number of pixels included in
       this analysis goes below 1000.  We conclude that in the $\ell$=30\deg
       field, about 55\% of the far-IR sources are mIRd, 20\% are mIRn,
       and 25\% are mIRb, while in the $\ell$=59\deg field, about 20\% are
       mIRd, 40\% are mIRn, and 40\% are mIRb.  It is evident from these
       plots, however, that these numbers are somewhat subjective and can
       vary depending on the cutoff selected. }
     \label{fig:percents}
     \end{figure}

   \begin{figure*}
     \centering
     \includegraphics[width=0.96\textwidth]{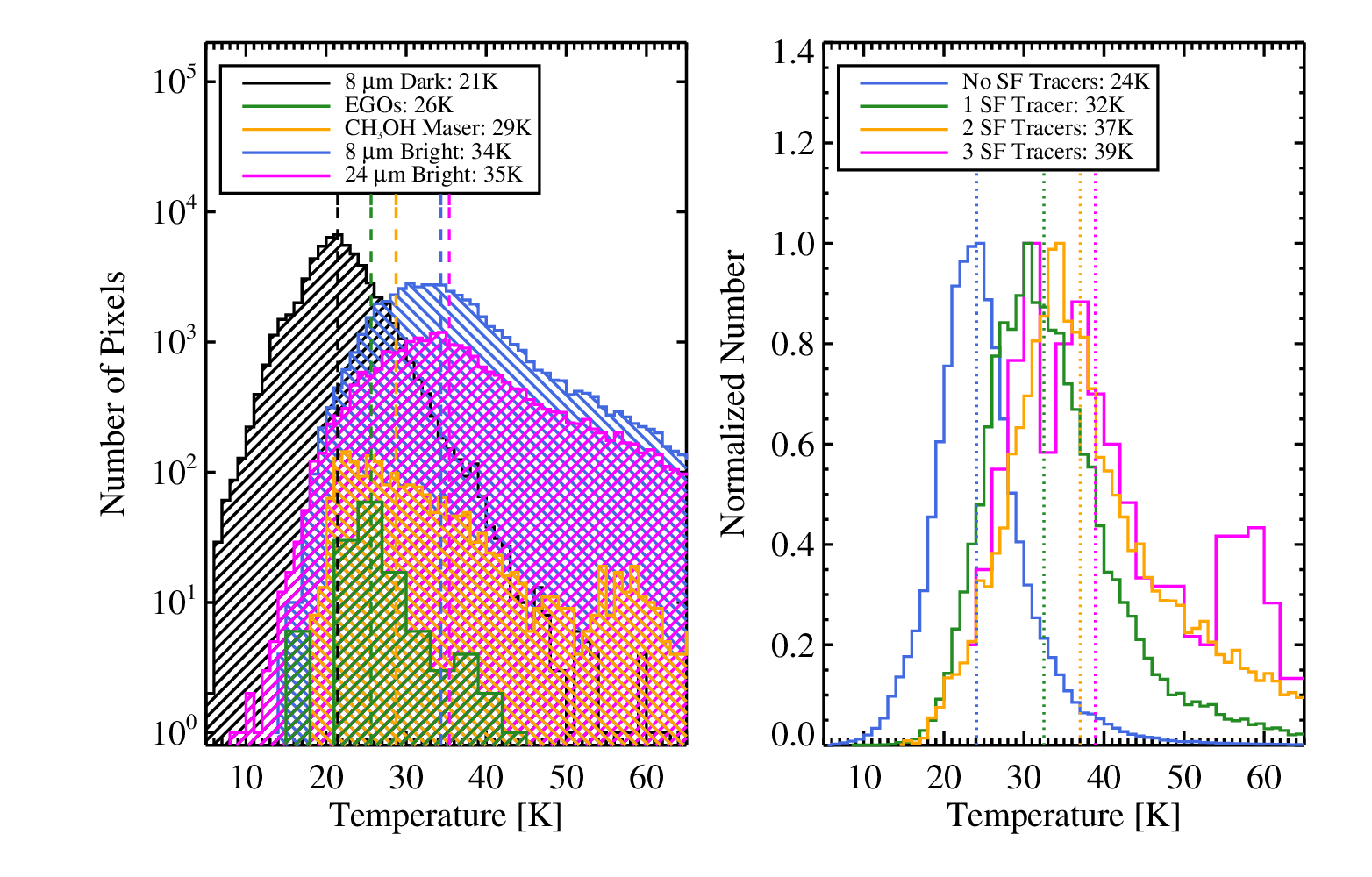}
     \caption{A comparison of star formation tracers in the $\ell$=30\deg field.
     The identification of these tracers is discussed in Section
     \ref{sec:sftracers}.  \textit{Left:} Log histogram of the temperature
     distributions for all source pixels identified as mIRd (8
     \micron~dark, black), associated with EGOs (green) or a CH$_{3}$OH
     maser (orange), mIRb (8 \micron~bright, blue), or 24 \micron~bright
     (magenta).  The median temperature is reported in the legend and with
     the dashed vertical line for each tracer.  If one assumes that a
     source will monotonically increase in temperature with time, this
     diagram gives a possible indication of the observed evolution of
     sources in this field.  \textit{Right:} A normalized temperature
     histogram for all pixels in the $\ell$=30\deg source masks containing 0
     (blue), 1 (green), 2 (orange), and 3 (magenta) of the star 
     formation tracers shown in the left panel.  The pixels containing 4
     star formation tracers are rare, follow the distribution of the EGOs,
     and have low number statistics and therefore are not plotted.  The median
     temperature is again reported in the legend and with dotted vertical
     lines.  This plot shows that warmer pixels are associated with more
     star formation tracers.}
     \label{fig:l30_histall}
   \end{figure*}

\section{Results}
\label{sec:res}
We discuss here the results of the modified blackbody fits, the properties
of the Galactic cirrus emission, a comparison of mIRb and mIRd pixels, and 
the association with star formation tracers.  We discuss what defines the
populations of mIRb and mIRd sources, and the objects which have no mid-IR
association.  Before the discussion of the results, we first mention the
many caveats and uncertainties that accompany these findings.  

\subsection{A Remark on Caveats and  Uncertainties}
\label{sec:caveats}
A major uncertainty in this analysis is the modified blackbody fits.
Firstly, star-forming regions show structure on many scales, and so fitting
a single column density and temperature to any point does not adequately
represent the whole region; rather it serves as a way to categorize
and differentiate between bulk, large-scale physical properties of
regions.  Additionally, this means that the quantities derived are likely
to be highly beam-diluted.  The true peak column density in
any given region is almost certainly higher than the beam-averaged column
density reported here.  Likewise, the beam-averaged temperatures reported
for cold objects are overestimates of the true temperature minimum, 
while the beam-averaged temperatures
reported for hot objects is an under-estimate of the peak.  These caveats explain
the somewhat small dynamic range observed in both temperature and column density. 

In addition, the parameters derived from the modified blackbody fits
themselves are rather uncertain.  For
most of the sources, the temperature is higher than 18 K, so all the points
used in the fit lie near the peak of the SED or on the Rayleigh-Jeans
slope.  This
equates to large uncertainties in the fits, especially at high temperatures. 
The median errors in the derived temperature are about 4 and 3  K (for the
$\ell$=30\deg and $\ell$=59\deg fields, respectively), while the median errors in the
derived column density are about 0.12 and 0.04 $\times$ 10$^{22}$ cm$^{-2}$
(in $\ell$=30\deg and $\ell$=59\deg fields, respectively).   The error is a strong
function of the parameter value, with a median temperature error of 13\%
and 18\% and median column density error of 44\% and 48\% (in the $\ell$=59\deg 
and $\ell$=30\deg field, respectively).
These uncertainties are
based on the 20\% calibration uncertainty assigned to the Hi-GAL fluxes.
Additionally, the derived quantities are highly dependent not only
on the background that is subtracted, but also on possible departures
between the gain scale of the PACS and the SPIRE data. Improvements to the
general issue of the Planck-Herschel cross calibration with respect to the accuracy of
the offset determination as well as to the gain values will be adressed in the future.

The assumption of a fixed $\beta$ can be considered to contribute the bulk
of the uncertainty in the temperature determination.  The degeneracy of
$\beta$-T has been discussed in \citet{par10}.  If we adopt a value
of $\beta$ = 1.5 (instead of 1.75), the temperatures are about 4K higher
on average and the column densities are about the same.  If we adopt a value 
of $\beta$ = 2.0 (instead of 1.75), the temperatures are about 3K lower on 
average, and again, the column densities are about the same.  With only
four data points to constrain the fits, fixing $\beta$ is nearly essential, 
with the choice of 1.75 being a reasonable guess for these
environments.  Variations in $\beta$ from 1.75 to 1.5 or 2 correspond to
changes in temperature of about 3 or 4 K.
There is the possibility that systematic changes in $\beta$, for
example a systematic decrease in $\beta$ toward the center of IRDCs due to
dust coagulation \citep{oss94}, could systematically change our results.
In the dark centers of IRDCs, however, a variation in $\beta$ only changes
the temperature by about 2 K (2 K hotter for $\beta$ = 1.5, and 2 K colder for
$\beta$ = 2.0), so while this is an important effect to consider, it is
unlikely to negate the observed trends in temperature.

The association with star formation tracers is also bridled with
uncertainty.  Each of the tracers included are not perfectly
represented with our methods.  The 8 and 24 \micron~label maps
are entirely dependent on the cutoffs chosen.  While we have done our best
to choose reasonable cutoffs, a
change in these cutoffs would have a significant effect on the properties inferred.
The association with EGOs suffers from a similar bias.  We have made a
careful attempt to identify probable EGO candidates, but the objects
themselves are somewhat ill-defined, so again, we resort to the by-eye
classification.  The association with CH$_{3}$OH
masers is robust in that it is based on uniform sensitivity blind surveys, 
however, these surveys suffer from large beam sizes and confusion which
consequently dilute their true physical properties.
All of these tracers also suffer from the inherent scatter that comes along with
varying levels of extinction and different distances.

Despite these caveats, the trends observed over many pixels remain robust.
The pixel-by-pixel comparison is an important technique for achieving the
best possible associations with the given data, and therefore, the most
robust statistics.  The object in the center of Figure
\ref{fig:filament_temp} highlights the importance of comparing values on a
pixel-by-pixel basis, rather than denoting the entire object as either
mIRb or mIRd.

\subsection{Properties of the Cirrus Cloud Emission}
While we primarily consider the cirrus cloud emission  as ``background'' in
this paper, we report here some useful physical properties derived in this analysis.
These maps were smoothed to 12' ($\Theta_{FWHM}$) resolution, so we report
only the average properties in Table \ref{table-cirrus}.  These data 
suggest that dust in diffuse clouds in
the molecular ring ($\ell$=30\deg field) is warmer, has higher column
densities, and a steeper spectral index ($\beta$) than dust in diffuse
clouds toward the mid-outer Galaxy, outside the molecular ring ($\ell$=59\deg field). 
In the cirrus background determination, we fit a Gaussian of the form:
\begin{math}g = B + A e^{{(b-b0)^{2}}/{2\sigma^{2}}}\end{math}
across Galactic latitude at each Galactic longitude.  We report here the
Gaussian fit to the median value across Galactic longitudes.  In the $\ell$=30\deg field, the
best fit parameters are A = 285 MJy/sr, B = 38 MJy/sr, b0
= -0.1\deg, and $\sigma$ = 0.4\deg, while in the $\ell$=59\deg field, the
best fit parameters are A = 53 MJy/sr, B = 29 MJy/sr, b0
= 0.1\deg, and $\sigma$ = 0.4\deg.  Previous studies \citep[e.g.][]{ros10}
have shown that the mean Galactic latitude where emission peaks depends strongly on
the Galactic latitude observed, but that the mean in the First Galactic
Quadrant is $<$b$>\approx$ -0.1\deg.  The $\ell$=30\deg field also shows an average
Galactic latitude of about -0.1\deg, while the $\ell$=59\deg field is higher
than the average at b0=0.1\deg.

\begin{table}
\caption{Smoothed Cirrus Emission Properties \label{table-cirrus}}
\centering
\begin{tabular}{cccc}
\hline\hline
Hi-GAL Field &
$\beta$\tablefootmark{1}
 & N(H$_{2}$)\tablefootmark{1} &
Temperature\tablefootmark{1} \\
   &   & ($\times$ 10$^{22}$ cm$^{-2})$ &
 (K) \\
\hline
$\ell$=30\deg  &  1.7 $\pm$ 0.2  &  0.7 $\pm$ 0.3  &  23 $\pm$ 1 \\

$\ell$=59\deg  &  1.5 $\pm$ 0.2  &  0.3 $\pm$ 0.1  &  21 $\pm$ 1 \\
\hline
\end{tabular}
\tablefoot{
  \tablefoottext{1}{The values given are the median of all valid values in the map,
  and the uncertainty quoted is the standard deviation.}}
\end{table}

\subsection{The Temperature and Column Density Maps}
We present full temperature and column density maps of the Hi-GAL $\ell$=30\deg
and $\ell$=59\deg SDP fields.  These maps are shown in Figure
\ref{fig:all_tempcol}  and are available to
download as FITS files online.  The source mask
label maps, the source mask temperature and column density maps, the error
maps of these quantities as well as
the star formation tracer label maps 
are all available
for download.  The maps presented in Figure \ref{fig:all_tempcol} are 
displayed such that the values inside the source masks represent the 
background-subtracted fit values, while the values outside the source mask 
are the fits to the background itself.

The uncertainties in these maps are discussed in Section
\ref{sec:caveats}.  In addition to those, we found that in the $\ell$=30\deg field,
our fits were returning unsensible parameters at high Galactic latitudes
($|b| \gtrsim$ 0.8\deg).  This is due to the imperfect calibration,
especially at these low levels, so the relative fluxes were no longer
physically meaningful.  We made several attempts at correcting or properly
ignoring those points, with little success, and therefore, recommend that
the maps in the $\ell$=30\deg field only be used within $|b| \leq$
0.8\deg.  Additionally, pixels near the edges of the source masks were just
barely above the background, and can produce unphysical fits, so be
cautious of any pixel that is right on the edge of the source mask.

The column density follows the far-IR flux closely, while the
temperature is quite varied, and in many cases, inversely correlated with
the column density.  The median temperatures for all the pixels in the source masks
are 26 and 20 K, while the median column densities are 0.25 and 0.10 
$\times$ 10$^{22}$ cm$^{-2}$ (for the $\ell$=30\deg and $\ell$=59\deg field,
respectively). 
The highest column density points are found in W43,
the large complex near $\ell$=30.75\deg, b=-0.05\deg \citep{bal10}, where 
the bright millimeter points, MM1 - MM4, have beam-averaged column
densities of around 10$^{23}$ cm$^{-2}$.  

\begin{table*}
\caption{Source Properties by Their Mid-IR Association\label{table-midIR}}
\centering
\begin{tabular}{clcccc}
\hline\hline
Hi-GAL &
Mid-IR & 
Temperature\tablefootmark{1} &
High Column Temperature\tablefootmark{2} &
Column Density\tablefootmark{1} &
Low Temperature Column\tablefootmark{3} \\
 Field & Classification &
 (K) &  (K) &  ($\times$ 10$^{22}$ cm$^{-2}$) 
& ($\times$ 10$^{22}$ cm$^{-2}$) \\        

\hline
$\ell$=30\deg  &  Dark (mIRd)    &  21   & 19  &  0.50 &  0.57 \\
          &  Neutral (mIRn) &  25   & 23  &  0.25 &  0.32 \\
          &  Bright (mIRb)  &  34   & 31  &  0.23 &  0.49 \\
$\ell$=59\deg  &  Dark (mIRd)    &  16   & 14  &  0.22 &  0.23 \\
          &  Neutral (mIRn) &  20   & 15  &  0.11 &  0.12 \\
          &  Bright (mIRb)  &  30   & 24  &  0.08 &  0.15 \\
\hline
\end{tabular}
\tablefoot{
  \tablefoottext{1}{The values given are the median of all valid values in the map.}
  \tablefoottext{2}{The median temperature of all valid points in the map
  with N(H$_{2}$) $>$ 10$^{22}$ cm$^{-2}$ in the $\ell$=30\deg field and 
  N(H$_{2}$) $>$ 0.3 $\times$ 10$^{22}$ cm$^{-2}$ in the $\ell$=59\deg field (to
  account for the fact that column densities in the $\ell$=59\deg field are
  about 1/3 of that in the $\ell$=30\deg field).}
  \tablefoottext{3}{The median column density of all valid points in the
  map with T $<$ 25 K. }
}
\end{table*}

\subsection{The association of Far-IR Clumps with mIRb and mIRd sources}
\label{sec:farirmir}
The association of Hi-GAL sources identified in the Far-IR with mid-IR
sources is of interest, as that gives some indication of their star-forming
activity.  We find that Hi-GAL sources span the range of pre- to
star-forming and that there exist significant trends in both
temperature and column density between these populations.
The growing interest in IRDCs as potential precursors to 
stellar clusters emphasizes the importance of identifying IRDC-like sources
throughout the Galaxy in an unbiased manner, not just where they exist in 
front of a bright
mid-IR background.  A survey such as Hi-GAL is essential to
identify IRDC-like objects throughout the Galaxy while a thorough analysis 
is necessary for understanding the trends 
in Hi-GAL sources that are pre- versus star-forming.

We find that mIRd pixels are characterized by colder temperatures (colder 
by more than 10 K) and higher column densities (about a factor of two 
higher) than mIRb pixels (see Table \ref{table-midIR}).  The mIRn pixels 
generally have a column density similar to that of mIRb pixels, and a 
temperature midway between mIRd and mIRn pixels (see Figure
\ref{fig:8micron_hist}).  This trend is also apparent in the left panel of 
Figure \ref{fig:l30_tempcol_sf}.  
Figure \ref{fig:8micron_hist} shows that introducing a column density
cutoff to the temperature distribution decreases the average temperature,
especially for mIRb pixels.  If, on the other hand, we introduce a
temperature cutoff (T $<$ 25 K) to the column density distributions, the
average column density increases.

A simple K-S test rules out the possibility that any combination of the
mIRd, mIRn, or mIRb populations in temperature or column density are
drawn from the same distribution with at least 99.7\% confidence.  For the
K-S test, we calculated the effective number of independent points in
each distribution by dividing the number of pixels in each distribution
by the number of pixels per beam.
Considering the errors discussed in Section
\ref{sec:caveats}, these trends are significant and not likely to be a
by-product of systematic errors.  Hi-GAL is sensitive to both cold, high
column density, likely pre-stellar sources and warmer, more diffuse star-forming
regions. We can distinguish between these using the temperature and column 
densities derived simply from Hi-GAL or through comparison with other
tracers. 

In Figure \ref{fig:percents}, we plot the fraction of mIRd, mIRn, and mIRb
pixels above the column density cutoff plotted on the x-axis.  In the
$\ell$=30\deg field, the mIRn fraction drops quickly as the mIRd fraction rises,
while in the $\ell$=59\deg field, mIRb fraction rises steadily as the mIRn
fraction drops.  While assigning an overall percentage to each of the
categories is somewhat subjective, because of the cutoff-dependent
variations, we estimate that in the $\ell$=30\deg field about 55\% of the Hi-GAL
identified source pixels are mIRd, 20\% are mIRn, and 25\% are mIRb.  In
the $\ell$=59\deg field, we estimate that about 20\% of the Hi-GAL identified
source pixels are mIRd, 40\% are mIRn, and 40\% are mIRb.  The fact that
the fraction of mIRd pixels in the $\ell$=59\deg field is so much lower than
that in the $\ell$=30\deg field could be an artifact of the relatively sparse 
mid-IR background emission in the $\ell$=59\deg field, or it could very well be
that there is a lower fraction of cold, high-column density clouds in the 
$\ell$=59\deg field.

\subsection{Star Formation Tracers in Hi-GAL Sources}
\label{sec:sftracerres}
We compare the temperature and column density maps with the star formation
tracer label maps (discussed in Section \ref{sec:sftracers}). 
While we evaluate the evolution of sources by their temperature alone,
\citet{beu10} analyze the evolution of four sources with changes over the
entire SED.
We find that the more star formation tracers associated with a source, the
higher the temperature (see Figure \ref{fig:l30_histall} right).  This
trend is not surprising: a star-forming clump will be warmer than a
pre-star-forming clump.  It is reasonable to assume that as a clump
evolves, the temperature will increase monotonically.  Therefore, we 
compare the individual star formation tracers with temperature to see if
there is any indication about which star formation tracer may turn on
first, and what the relative lifetimes might be. 
There are a lot of uncertainties in this analysis (discussed in
\ref{sec:caveats}) and we are comparing beam-averaged physical parameters, 
so there is quite a lot of scatter.

In the left panel of Figure \ref{fig:l30_histall} we see a progression of 
star formation tracers with temperature, with a huge amount of overlap.  
The median temperatures of each population (denoted with a dotted line, 
and stated in the legend) delineate a sequence from cold pre-stellar clumps to
warm star-forming clumps, from mIRd, to outflow/maser sources, to sources
which are bright at 8 and 24 \micron.  However, the overall distributions are very
wide, possibly indicating long lifetimes, but more likely an indication of
large errors in both the beam-averaged physical properties and the
assignment of star formation tracers.  

Figure \ref{fig:l30_tempcol_sf} shows these trends in both temperature and
column density space.  The left panel shows that mIRd pixels are generally
colder and have higher column densities than mIRb pixels.  The right panel
shows that all the star formation tracer pixels are warmer than the mIRd
population, with CH$_{3}$OH masers in the mid-range of temperatures, and 24
and 8 \micron~bright pixels occupying the highest temperature range.  EGOs
are characterized by a similar temperature as CH$_{3}$OH masers, but a
slightly higher column density.  Interestingly, the EGOs seem to occupy 
two distinct regions on the plot in Figure \ref{fig:l30_tempcol_sf}, one
higher and one lower column density, indicating that EGOs are present in
a variety of environments.  It is not surprising that the EGOs
represent the highest column density population, as they are the most
localized star formation tracer we have used.  

We perform simple K-S tests to determine the likelihood that any
combination of the temperature or column density distributions of 
star formation tracers (pixels which are mIRd, mIRb, 24 \micron~bright,
or contain EGOs or masers) are drawn from the same distribution.  For the
K-S test, we calculated the effective number of independent points in
each distribution by dividing the number of pixels in each distribution
by the number of pixels per beam (about 23 in the 25" resolution images used
for all tracers except the masers, which have an accuracy of about 33",
or 41 pixels).  Since EGOs are defined locally, each pixel is considered 
independent.  When we compare the temperature distributions of all
combinations of the mIRd, mIRb, 24 \micron~bright, EGO and maser
populations, we can rule out with at least 99.7\% confidence that any are
drawn from the same distribution, except the 24 and 8 \micron~bright points, 
which have about a 55\% probability of being drawn from the same
distribution, and the EGO and maser populations which have about a 0.5\%
chance of being drawn from the same distribution.  When we compare the
column density distributions of all combinations of the mIRd, mIRb, 24 
\micron~bright, EGO and maser populations, we can rule at with at least 
99.7\% confidence that any are drawn from the same distribution, except the
EGO and maser populations, which have about a 55\% chance of being drawn
from the same distribution.  The fact that EGOs and CH$_{3}$OH masers may
be tracing the same population of sources is encouraging, as they are both
supposed to trace young, massive outflows.  We would also have expected that
pixels which are bright at 8 or 24 \micron~should be tracing similar
environments, as both can be excited by the heat or UV light from a young,
accreting protostar.

We see interesting trends in Hi-GAL clumps between temperature and column
density and star formation tracers.  We find that 8 \micron~dark pixels are, on
average, the coldest, followed by pixels containting EGOs and/or CH$_{3}$OH
masers, then 8 and 24 \micron~bright points.  If we assume
that as a clump evolves, it will monotonically increase in temperature with
time, then this could suggest a possible sequence of tracers.  However, the
scatter is still too large to suggest a definitive sequence or lifetimes. 
We should expect that one of the first stages will be the
formation of an outflow, detectable by an EGO or CH$_{3}$OH maser.
Following that, the protostar will continue to heat its surroundings 
and light up at 24
\micron.  A massive protostar may form an UCHII region while still
accreting, whose UV light would excite PAHs in the 8 \micron~band of
GLIMPSE.  Unlike \citet{bat10}, we find that 24 \micron~ sources light up
around the same time as 8 \micron~sources.  This is almost certainly due to
the use of a high, automated cutoff to determine the existence or absence
of a 24 \micron~source.  In \citet{bat10}, UCHII regions were associated
with very bright ($\gtrsim$ 1 Jy) 24 \micron~point sources, which may be the
population of bright sources we are picking out here.  We note that
even though most or all UCHII regions are mIRb, certainly not all mIRb
sources are UCHII regions \citep[e.g.][]{mot10}.  The fact that the
population of mIRb and 24 \micron~bright are so similar is likely an
artifact of the cutoffs chosen.  The two
should be associated for bright sources, which is the trend we see in this
paper, but for dimmer sources, which are excluded by the cutoff, we do not
know the association.  The detection of an EGO or 24
\micron~point source depends on the exinction and the background, and the
sensitivity for detection will decrease with distance.
As discussed, there are many caveats and systematic errors
to consider, however, the trends are intriguing.  

    \begin{figure}
     \centering
       \includegraphics
	   [width=0.46\textwidth]{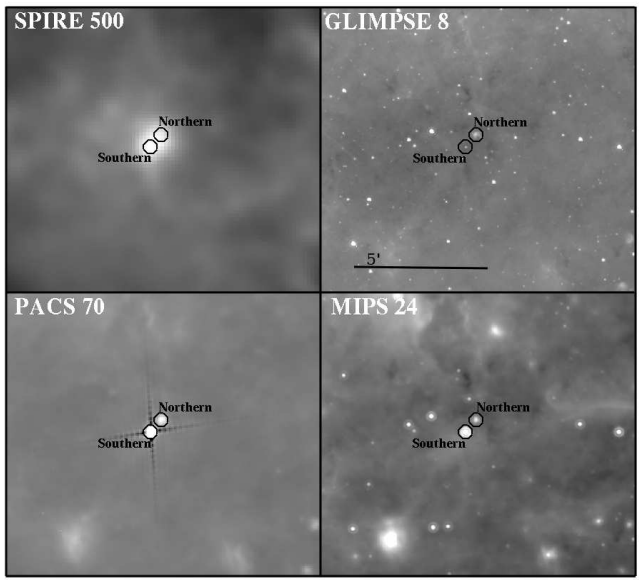}\\
     \caption{A possible IRDC-like source on the far-side of the Galaxy.
     This source is cold ($\sim$ 20 K), has high column density ($\sim 3 
     \times 10^{22}$ cm$^{-2}$) and is amidst the bright mid-IR Galactic
     Plane at ($\ell$,b) = (30.01, -0.27).  By these measures, it should be an
     IRDC, but the 8 \micron~image (top right) shows only a moderate decrement
     and the near-side extinction column density is almost an order of magnitude
     less than expected from the far-IR column density.  This source breaks
     up into a Northern ($\ell$,b) = (30.003,-0.266)and Southern ($\ell$,b) =
     (30.009,-0.274) component at the higher resolution of the PACS 70 
     \micron~band.  This source is a great candidate for further study.}
     \label{fig:farsideirdc}
     \end{figure}

\begin{table*}
\caption{Far-Side IRDC Candidates \label{table-irdclike}}
\centering
\begin{tabular}{lccccc}
\hline\hline
  &
  &  &
Hi-GAL & Extinction\tablefootmark{1} &
  \\
 &
l & b &
Column & Column &
Temperature \\
Source Name  &  ($^{\circ}$) & ($^{\circ}$) &
($\times 10^{22} \rm{cm}^{-2}$) &
($\times 10^{22} \rm{cm}^{-2}$) &
(K) \\
\hline
G030.60+0.18  &  30.602	 &  +0.177  &  6.2    &  1.1     &  22  \\
G030.34-0.11  &  30.345	 &  -0.114  &  1.7  &  0.15  &	19  \\	
G030.01-0.27\tablefootmark{2}  &  30.006	 &  -0.270  &  3.1  &  0.5   & 	21  \\	
G029.31-0.05  &  29.312	 &  -0.051  &  1.0  &  0.06  &	13  \\
G029.28-0.33  &  29.284	 &  -0.331  &  1.6  &  0.02   &  20  \\
\hline
\end{tabular}
\tablefoot{
  \tablefoottext{1}{Not knowing the distances to all these sources, we use
  the \citet{per09} extinction column density estimation method, which
  basically assumes the cloud is on the near-side of the Galaxy}
  \tablefoottext{2}{This is the source that is discussed in the text in
  Section \ref{sec:farirdc}.}
}
\end{table*}
 
\section{IRDC-like Source Candidates on the Far-Side of the Galaxy}
\label{sec:farirdc}
We present a list of candidate IRDCs on the far-side of the Galaxy in Table
\ref{table-irdclike}.  These sources are cold ($\sim$ 20 K), high column
density ($>$ 1 $\times$ 10$^{22}$ cm$^{-2}$) objects identified in
Hi-GAL with very weak absorption at 8 \micron.  These are objects that,
were they in front of the bright mid-IR background, \textit{should} be
IRDCs, but are not.  To identify these sources,
we calculated an extinction column density map at 8 \micron, using the
method of \citet{per09} and compared this with the Hi-GAL column density
map.  Any sources with a Hi-GAL column density $>$ 1 $\times$ 10$^{22}$
cm$^{-2}$ and an extinction column density several (3-4) factors lower was
considered.  We performed an examination by eye simply to check that these are
plausible candidates.  The extinction must be weak, the region cannot be
mIRb, and it needs to be cold ($<$ 30 K).  In other words, these are
sources that, if on the near-side of the Galaxy, should be dark at 8
\micron, yet they are not.  None of these sources are identified as IRDCs
by the catalog of \citet{per09}.  Sources at high Galactic
latitudes ($|b| \gtrsim 0.5$\deg) and in the $\ell$=59\deg field were not
included because the sparse background adds additional uncertainty to the
extinction column density estimate.

We discuss G030.01-0.27, a candidate far-side IRDC, in more detail here as
an example of this class of objects.  G030.01-0.27 is a high-column density 
($\sim$ 3 $\times$ 10$^{22}$
cm$^{-2}$), cold ($\sim$ 20 K) dust continuum source in the $\ell$=30\deg 
field near the Galactic mid-plane that is neither 8 \micron~dark nor
bright (see Figure \ref{fig:farsideirdc}).  There is a slight decrement at 
8 \micron~that is nearly consistent
with background fluctuations as this source was not detected in the IRDC
catalogs of \citet{per09} or \citet{sim06}. 
 At 70 \micron, this source breaks up into two, a Southern source at 
($\ell$,b) = (30.009,-0.274) and Northern source at ($\ell$,b) = (30.003,-0.266).  
The Southern source appears to be very small and faint at 8 \micron, while 
the Northern source is a faint, slightly extended source at 8 \micron.
The Southern source has bright 24 and 70 \micron~emission.  The Northern
source is also seen in emission at these wavelengths, though less
brightly.  Comparison with the MAGPIS \citep{hel06,whi05} 20 cm maps show no 
emission in the radio continuum coincident with either source, while the
MAGPIS 6 cm map shows a very faint object associated with the Southern
source.

The faint decrement at 8 \micron~can be morphologically matched 
with a molecular cloud in $^{13}$CO \citep[][from the
  Galactic Ring Survey]{jac06} at about 104 kms$^{-1}$.  Integrating from
100 to 108 kms$^{-1}$, assuming 20 K, and using Eq. 17 from \citet{bat10}
the column density is about 2.5 $\times 10^{22}$ cm$^{-2}$, consistent with
the Hi-GAL column density estimate.  Using the rotation curve of
\citet{rei09}, the near-side distance of this $^{13}$CO cloud is 5.8 kpc
and the far-side distance is 8.7 kpc.  Using the extinction mass
determination technique of \citet{but09} as applied in \citet{bat10}, the
8 \micron~extinction column density of the cloud is 0.5 $\times 10^{22}$
cm$^{-2}$ assuming it is on the near-side of the Galaxy or 1 $\times 10^{22}$
cm$^{-2}$ assuming it is on the far-side of the Galaxy.  It is not expected
that the extinction column density estimation method will be very robust
for clouds on the far-side of the Galaxy, but this very slight extinction
is at least consistent with the cloud being on the far-side of the Galaxy.
Additionally, this cloud shows no obvious HI self-absorption \citep{sti06}.
While none of this evidence is conclusive, it all seems to indicate that 
this cloud is on the far-side of the Galaxy.  Another possibility, however,
would be that this cloud is simply \textit{behind} the majority of the
diffuse mid-IR emission, and still on the near-side of the Galaxy.  Either
way, it is an interesting source that was otherwise missed as such in
previous study.

The fact that the mid-IR background is bright at the locations of these
sources, yet the 8 \micron~extinction column density is extremely low
is a strong indication that these sources are \textit{behind} the
bright mid-IR background.  In fact, the Hi-GAL distance analysis by
\citet{rus11} finds that all the sources in Table \ref{table-irdclike} are
either at the far or tangent distance ($>$ 7 kpc) except for
G029.28-0.33, which is on the near-side at about 6 kpc, so still likely behind
most of the bright mid-IR emission.

With their high far-IR column density and low
temperature, these sources are easily categorized (based on the left panel of Figure 
\ref{fig:l30_tempcol_sf}) as IRDC-like.  These are the first
candidate far-side IRDCs of which the authors are aware.  The
potential outer Galaxy IRDC identified by \citet{fri07} is similar in that
they are both IRDC-like objects identified independent of 8
\micron~absorption.  However, these sources are not in the outer Galaxy, but
on the far-side of the inner Galaxy.  These candidates are just a few of
potentially many more IRDC-like sources which remain to be uncovered in a
study like this over the Galactic Plane.

\section{Conclusions}
\label{sec:conc}
We have performed cirrus-subtracted pixel-by-pixel modified
blackbody fits to the Hi-GAL $\ell$=30\deg and $\ell$=59\deg SDP fields.  The source
identification and cirrus-subtraction routines are robust and can be
applied to Hi-GAL data throughout the Galactic Plane.  We present
temperature and column density maps of the dense clumps in these fields and
cirrus column density, temperature, and $\beta$ maps.
We find that the cirrus cloud emission is
characterized by $\beta$ = 1.7, N(H$_{2}) = 0.7 \times 10^{22} \rm{cm}^{-2}$,
and T = 23 K in the $\ell$=30\deg field and $\beta$ = 1.5, N(H$_{2}) = 0.3 
\times 10^{22} \rm{cm}^{-2}$, and T = 21 K in the $\ell$=59\deg field.

We also characterize
each pixel as mid-IR-bright (mIRb), mid-IR-dark (mIRd), or mid-IR-neutral
(mIRn), based on the contrast at 8 \micron.  
The association of Hi-GAL sources identified in the far-IR with mid-IR
sources is of interest, as the far-IR sources span the range of pre- to
star-forming regions, and the mid-IR can help to separate these.  We find 
that in the $\ell$=30\deg
field, about 55\% of the pixels are mIRd, 20\%  are mIRn, and 25\% are
mIRb, while in the $\ell$=59\deg field, about 20\% are mIRd, 40\% are mIRn, and 
40\% are mIRb.  There exist significant trends in temperature and column
density between the populations of mIRd to mIRb.  We find that mIRd dark 
pixels are about 10 K colder
and a factor of two higher column density than mIRb pixels.  The mIRd
pixels are likely cold pre-star-forming regions, while the mIRb pixels are
in regions that have probably begun to form stars.  This study has shown
that Hi-GAL-identified sources span the range from cold, pre-star-forming
to actively star-forming regions.  

We also include the presence or absence of EGOs (Extended Green Objects,
indicative of shocks in outflows), CH$_{3}$OH masers, and emission at 8 and
24 \micron~as star formation tracers.  We find that warmer pixels are
associated with more star formation tracers.  We also
find a wide but plausible trend in temperature, where the coldest pixels,
on average, are mid-IR-dark, followed by pixels containing EGOs and/or
CH$_{3}$OH masers, then 8 and 24 \micron~bright sources.
While the systematic errors are too large to suggest an
evolutionary sequence, this trend is intriguing.

Finally, we identify five candidate far-side IRDCs.  These objects are cold
($\sim$ 20 K), high column density ($>$ 1 $\times$ 10$^{22}$ cm$^{-2}$)
sources identified with Hi-GAL that have very weak or no absorption at 8
\micron.  We explore one such candidate in more detail in Section
\ref{sec:farirdc}. 
This object at roughly
($\ell$,b) = (30.01, -0.27) has a high far-IR column density
(N(H$_{2}) \sim 3 \times 10^{22} \rm{cm}^{-2}$), is cold ($\sim$ 20 K) and in a
position near an abundantly bright mid-IR background, yet shows almost no 
decrement at 8 \micron.  In
fact, the 8 \micron~extintion-derived column density is almost an order of
magnitude lower than expected from the far-IR estimate.
These candidate far-side IRDCs are the first of their kind of which the
authors are aware.
This type of analysis will likely uncover many more such
objects.  With a complete sample of IRDC-like (cold, high column density) 
clouds, independent of the local mid-IR background,
one could map the clouds in the earliest stages of star-formation over the
entire Galaxy.  

\begin{acknowledgements}
Data processing and map production has been possible
thanks to generous support from the Italian Space Agency via contract
I/038/080/0. Data presented in this paper were also analyzed
using The Herschel interactive processing environment (HIPE), a joint development
by the Herschel Science Ground Segment Consortium, consisting of ESA,
the NASA Herschel Science Center, and the HIFI, PACS, and SPIRE consortia.
This research made use of Montage, funded by the National Aeronautics and
Space Administration's Earth Science Technology Office, Computation
Technologies Project, under Cooperative Agreement Number NCC5-626 between 
NASA and the California Institute of Technology. Montage is maintained by 
the NASA/IPAC Infrared Science Archive.  This work made use of ds9, the
Goddard Space Flight Center's IDL Astronomy Library, and the NASA
Astrophysics Data System Bibliographic Services.
C.B. is supported by the National Science
Foundation (NSF) through the Graduate Research Fellowship Program (GRFP).

\end{acknowledgements}

\bibliographystyle{aa}
\bibliography{references1}

\begin{thebibliography}{61}
\expandafter\ifx\csname natexlab\endcsname\relax\def\natexlab#1{#1}\fi

\bibitem[{{Aguirre} {et~al.}(2010){Aguirre}, {Ginsburg}, {Dunham}, {Drosback},
  {Bally}, {Battersby}, {Bradley}, {Cyganowski}, {Dowell}, {Evans}, {Glenn},
  {Harvey}, {Rosolowsky}, {Stringfellow}, {Walawender}, \& {Williams}}]{agu10}
{Aguirre}, J.~E., {Ginsburg}, A.~G., {Dunham}, M.~K., {et~al.} 2010, ArXiv
  e-prints

\bibitem[{{Bally} {et~al.}(2010){Bally}, {Anderson}, {Battersby}, {Calzoletti},
  {Digiorgio}, {Faustini}, {Ginsburg}, {Li}, {Nguyen-Luong}, {Molinari},
  {Motte}, {Pestalozzi}, {Plume}, {Rodon}, {Schilke}, {Schlingman},
  {Schneider-Bontemps}, {Shirley}, {Stringfellow}, {Testi}, {Traficante},
  {Veneziani}, \& {Zavagno}}]{bal10}
{Bally}, J., {Anderson}, L.~D., {Battersby}, C., {et~al.} 2010, \aap, 518, L90+

\bibitem[{{Battersby} {et~al.}(2010){Battersby}, {Bally}, {Jackson},
  {Ginsburg}, {Shirley}, {Schlingman}, \& {Glenn}}]{bat10}
{Battersby}, C., {Bally}, J., {Jackson}, J.~M., {et~al.} 2010, \apj, 721, 222

\bibitem[{{Benjamin} {et~al.}(2003){Benjamin}, {Churchwell}, {Babler}, {Bania},
  {Clemens}, {Cohen}, {Dickey}, {Indebetouw}, {Jackson}, {Kobulnicky},
  {Lazarian}, {Marston}, {Mathis}, {Meade}, {Seager}, {Stolovy}, {Watson},
  {Whitney}, {Wolff}, \& {Wolfire}}]{ben03}
{Benjamin}, R.~A., {Churchwell}, E., {Babler}, B.~L., {et~al.} 2003, \pasp,
  115, 953

\bibitem[{{Bernard} {et~al.}(2010){Bernard}, {Paradis}, {Marshall}, {Montier},
  {Lagache}, {Paladini}, {Veneziani}, {Brunt}, {Mottram}, {Martin},
  {Ristorcelli}, {Noriega-Crespo}, {Compi{\`e}gne}, {Flagey}, {Anderson},
  {Popescu}, {Tuffs}, {Reach}, {White}, {Benedetti}, {Calzoletti}, {Digiorgio},
  {Faustini}, {Juvela}, {Joblin}, {Joncas}, {Mivilles-Deschenes}, {Olmi},
  {Traficante}, {Piacentini}, {Zavagno}, \& {Molinari}}]{Ber10}
{Bernard}, J., {Paradis}, D., {Marshall}, D.~J., {et~al.} 2010, \aap, 518, L88+

\bibitem[{{Beuther} {et~al.}(2010){Beuther}, {Henning}, {Linz}, {Krause},
  {Nielbock}, \& {Steinacker}}]{beu10}
{Beuther}, H., {Henning}, T., {Linz}, H., {et~al.} 2010, \aap, 518, L78+

\bibitem[{{Beuther} \& {Sridharan}(2007)}]{beu07b}
{Beuther}, H. \& {Sridharan}, T.~K. 2007, \apj, 668, 348

\bibitem[{{Beuther} {et~al.}(2007){Beuther}, {Zhang}, {Bergin}, {Sridharan},
  {Hunter}, \& {Leurini}}]{beu07a}
{Beuther}, H., {Zhang}, Q., {Bergin}, E.~A., {et~al.} 2007, \aap, 468, 1045

\bibitem[{{Bressert} {et~al.}(2010){Bressert}, {Bastian}, {Gutermuth},
  {Megeath}, {Allen}, {Evans}, {Rebull}, {Hatchell}, {Johnstone}, {Bourke},
  {Cieza}, {Harvey}, {Merin}, {Ray}, \& {Tothill}}]{bre11}
{Bressert}, E., {Bastian}, N., {Gutermuth}, R., {et~al.} 2010, \mnras, 409, L54

\bibitem[{{Butler} \& {Tan}(2009)}]{but09}
{Butler}, M.~J. \& {Tan}, J.~C. 2009, \apj, 696, 484

\bibitem[{{Carey} {et~al.}(1998){Carey}, {Clark}, {Egan}, {Price}, {Shipman},
  \& {Kuchar}}]{car98}
{Carey}, S.~J., {Clark}, F.~O., {Egan}, M.~P., {et~al.} 1998, \apj, 508, 721

\bibitem[{{Carey} {et~al.}(2009){Carey}, {Noriega-Crespo}, {Mizuno}, {Shenoy},
  {Paladini}, {Kraemer}, {Price}, {Flagey}, {Ryan}, {Ingalls}, {Kuchar},
  {Pinheiro Gon{\c c}alves}, {Indebetouw}, {Billot}, {Marleau}, {Padgett},
  {Rebull}, {Bressert}, {Ali}, {Molinari}, {Martin}, {Berriman}, {Boulanger},
  {Latter}, {Miville-Deschenes}, {Shipman}, \& {Testi}}]{car09}
{Carey}, S.~J., {Noriega-Crespo}, A., {Mizuno}, D.~R., {et~al.} 2009, \pasp,
  121, 76

\bibitem[{{Chambers} {et~al.}(2009){Chambers}, {Jackson}, {Rathborne}, \&
  {Simon}}]{cha09}
{Chambers}, E.~T., {Jackson}, J.~M., {Rathborne}, J.~M., \& {Simon}, R. 2009,
  \apjs, 181, 360

\bibitem[{{Compi{\`e}gne} {et~al.}(2010){Compi{\`e}gne}, {Flagey},
  {Noriega-Crespo}, {Martin}, {Bernard}, {Paladini}, \& {Molinari}}]{Com10}
{Compi{\`e}gne}, M., {Flagey}, N., {Noriega-Crespo}, A., {et~al.} 2010, \apjl,
  724, L44

\bibitem[{{Cyganowski} {et~al.}(2009){Cyganowski}, {Brogan}, {Hunter}, \&
  {Churchwell}}]{cyg09}
{Cyganowski}, C.~J., {Brogan}, C.~L., {Hunter}, T.~R., \& {Churchwell}, E.
  2009, ArXiv e-prints

\bibitem[{{Cyganowski} {et~al.}(2008){Cyganowski}, {Whitney}, {Holden},
  {Braden}, {Brogan}, {Churchwell}, {Indebetouw}, {Watson}, {Babler},
  {Benjamin}, {Gomez}, {Meade}, {Povich}, {Robitaille}, \& {Watson}}]{cyg08}
{Cyganowski}, C.~J., {Whitney}, B.~A., {Holden}, E., {et~al.} 2008, \aj, 136,
  2391

\bibitem[{{de Wit} {et~al.}(2005){de Wit}, {Testi}, {Palla}, \&
  {Zinnecker}}]{dew05}
{de Wit}, W.~J., {Testi}, L., {Palla}, F., \& {Zinnecker}, H. 2005, \aap, 437,
  247

\bibitem[{{Desert} {et~al.}(1990){Desert}, {Boulanger}, \& {Puget}}]{des90}
{Desert}, F., {Boulanger}, F., \& {Puget}, J.~L. 1990, \aap, 237, 215

\bibitem[{{Egan} {et~al.}(1998){Egan}, {Shipman}, {Price}, {Carey}, {Clark}, \&
  {Cohen}}]{ega98}
{Egan}, M.~P., {Shipman}, R.~F., {Price}, S.~D., {et~al.} 1998, \apjl, 494,
  L199+

\bibitem[{{Ellingsen}(1996)}]{ell96thesis}
{Ellingsen}, S. 1996, PhD thesis, Physics Department, University of Tasmania,
  GPO Box 252C, Hobart 7001, Australia;
  <EMAIL>Ellingsen@phys.utas.edu.au</EMAIL>

\bibitem[{{Ellingsen} {et~al.}(1996){Ellingsen}, {von Bibra}, {McCulloch},
  {Norris}, {Deshpande}, \& {Phillips}}]{ell96}
{Ellingsen}, S.~P., {von Bibra}, M.~L., {McCulloch}, P.~M., {et~al.} 1996,
  \mnras, 280, 378

\bibitem[{{Frieswijk} {et~al.}(2007){Frieswijk}, {Spaans}, {Shipman},
  {Teyssier}, \& {Hily-Blant}}]{fri07}
{Frieswijk}, W.~W.~F., {Spaans}, M., {Shipman}, R.~F., {Teyssier}, D., \&
  {Hily-Blant}, P. 2007, \aap, 475, 263

\bibitem[{{Gautier} {et~al.}(1992){Gautier}, {Boulanger}, {Perault}, \&
  {Puget}}]{gau92}
{Gautier}, III, T.~N., {Boulanger}, F., {Perault}, M., \& {Puget}, J.~L. 1992,
  \aj, 103, 1313

\bibitem[{{Gieles} \& {Portegies Zwart}(2011)}]{gie11}
{Gieles}, M. \& {Portegies Zwart}, S.~F. 2011, \mnras, 410, L6

\bibitem[{{Griffin} {et~al.}(2010){Griffin}, {Abergel}, {Abreu}, {Ade},
  {Andr?e}, {Augueres}, {Babbedge}, {Bae}, {Baillie}, {Baluteau}, {Barlow},
  {Bendo}, {Benielli}, {Bock}, {Bonhomme}, {Brisbin}, {Brockley-Blatt},
  {Caldwell}, {Cara}, {Castro-Rodriguez}, {Cerulli}, {Chanial}, {Chen},
  {Clark}, {Clements}, {Clerc}, {Coker}, {Communal}, {Conversi}, {Cox},
  {Crumb}, {Cunningham}, {Daly}, {Davis}, {De Antoni}, {Delderfield}, {Devin},
  {Di Giorgio}, {Didschuns}, {Dohlen}, {Donati}, {Dowell}, {Dowell}, {Duband},
  {Dumaye}, {Emery}, {Ferlet}, {Ferrand}, {Fontignie}, {Fox}, {Franceschini},
  {Frerking}, {Fulton}, {Garcia}, {Gastaud}, {Gear}, {Glenn}, {Goizel},
  {Griffin}, {Grundy}, {Guest}, {Guillemet}, {Hargrave}, {Harwit}, {Hastings},
  {Hatziminaoglou}, {Herman}, {Hinde}, {Hristov}, {Huang}, {Imhof}, {Isaak},
  {Israelsson}, {Ivison}, {Jennings}, {Kiernan}, {King}, {Lange}, {Latter},
  {Laurent}, {Laurent}, {Leeks}, {Lellouch}, {Levenson}, {Li}, {Li},
  {Lilienthal}, {Lim}, {Liu}, {Lu}, {Madden}, {Mainetti}, {Marliani}, {McKay},
  {Mercier}, {Molinari}, {Morris}, {Moseley}, {Mulder}, {Mur}, {Naylor},
  {Nguyen}, {O'Halloran}, {Oliver}, {Olofsson}, {Olofsson}, {Orfei}, {Page},
  {Pain}, {Panuzzo}, {Papageorgiou}, {Parks}, {Parr-Burman}, {Pearce},
  {Pearson}, {P?erez-Fournon}, {Pinsard}, {Pisano}, {Podosek}, {Pohlen},
  {Polehampton}, {Pouliquen}, {Rigopoulou}, {Rizzo}, {Roseboom}, {Roussel},
  {Rowan-Robinson}, {Rownd}, {Saraceno}, {Sauvage}, {Savage}, {Savini},
  {Sawyer}, {Scharmberg}, {Schmitt}, {Schneider}, {Schulz}, {Schwartz},
  {Shafer}, {Shupe}, {Sibthorpe}, {Sidher}, {Smith}, {Smith}, {Smith},
  {Spencer}, {Stobie}, {Sudiwala}, {Sukhatme}, {Surace}, {Stevens}, {Swinyard},
  {Trichas}, {Tourette}, {Triou}, {Tseng}, {Tucker}, {Turner}, {Vaccari},
  {Valtchanov}, {Vigroux}, {Virique}, {Voellmer}, {Walker}, {Ward}, {Waskett},
  {Weilert}, {Wesson}, {White}, {Whitehouse}, {Wilson}, {Winter}, {Woodcraft},
  {Wright}, {Xu}, {Zavagno}, {Zemcov}, {Zhang}, \& {Zonca}}]{gri10}
{Griffin}, M.~J., {Abergel}, A., {Abreu}, A., {et~al.} 2010, ArXiv e-prints

\bibitem[{{Helfand} {et~al.}(2006){Helfand}, {Becker}, {White}, {Fallon}, \&
  {Tuttle}}]{hel06}
{Helfand}, D.~J., {Becker}, R.~H., {White}, R.~L., {Fallon}, A., \& {Tuttle},
  S. 2006, \aj, 131, 2525

\bibitem[{{Jackson} {et~al.}(2006){Jackson}, {Rathborne}, {Shah}, {Simon},
  {Bania}, {Clemens}, {Chambers}, {Johnson}, {Dormody}, {Lavoie}, \&
  {Heyer}}]{jac06}
{Jackson}, J.~M., {Rathborne}, J.~M., {Shah}, R.~Y., {et~al.} 2006, \apjs, 163,
  145

\bibitem[{{Kauffmann} {et~al.}(2008){Kauffmann}, {Bertoldi}, {Bourke}, {Evans},
  \& {Lee}}]{kau08}
{Kauffmann}, J., {Bertoldi}, F., {Bourke}, T.~L., {Evans}, II, N.~J., \& {Lee},
  C.~W. 2008, \aap, 487, 993

\bibitem[{{Kauffmann} \& {Pillai}(2010)}]{kau10}
{Kauffmann}, J. \& {Pillai}, T. 2010, \apjl, 723, L7

\bibitem[{{Lada} \& {Lada}(2003)}]{lad03}
{Lada}, C.~J. \& {Lada}, E.~A. 2003, \araa, 41, 57

\bibitem[{{Markwardt}(2009)}]{mar09}
{Markwardt}, C.~B. 2009, in Astronomical Society of the Pacific Conference
  Series, Vol. 411, Astronomical Society of the Pacific Conference Series, ed.
  {D.~A.~Bohlender, D.~Durand, \& P.~Dowler}, 251--+

\bibitem[{{Martin} {et~al.}(2010){Martin}, {Miville-Desch{\^e}nes}, {Roy},
  {Bernard}, {Molinari}, {Billot}, {Brunt}, {Calzoletti}, {Digiorgio}, {Elia},
  {Faustini}, {Joncas}, {Mottram}, {Natoli}, {Noriega-Crespo}, {Paladini},
  {Robitaille}, {Strafella}, {Traficante}, \& {Veneziani}}]{mar10}
{Martin}, P.~G., {Miville-Desch{\^e}nes}, M., {Roy}, A., {et~al.} 2010, \aap,
  518, L105+

\bibitem[{{Minier} {et~al.}(2001){Minier}, {Conway}, \& {Booth}}]{min01}
{Minier}, V., {Conway}, J.~E., \& {Booth}, R.~S. 2001, \aap, 369, 278

\bibitem[{{Minier} {et~al.}(2003){Minier}, {Ellingsen}, {Norris}, \&
  {Booth}}]{min03}
{Minier}, V., {Ellingsen}, S.~P., {Norris}, R.~P., \& {Booth}, R.~S. 2003,
  \aap, 403, 1095

\bibitem[{{Miville-Desch{\^e}nes} \& {Lagache}(2005)}]{Miv05}
{Miville-Desch{\^e}nes}, M. \& {Lagache}, G. 2005, \apjs, 157, 302

\bibitem[{{Molinari} {et~al.}(2010{\natexlab{a}}){Molinari}, {Swinyard},
  {Bally}, {Barlow}, {Bernard}, {Martin}, {Moore}, {Noriega-Crespo}, {Plume},
  {Testi}, {Zavagno}, {Abergel}, {Ali}, {Andr{\'e}}, {Baluteau}, {Benedettini},
  {Bern{\'e}}, {Billot}, {Blommaert}, {Bontemps}, {Boulanger}, {Brand},
  {Brunt}, {Burton}, {Campeggio}, {Carey}, {Caselli}, {Cesaroni}, {Cernicharo},
  {Chakrabarti}, {Chrysostomou}, {Codella}, {Cohen}, {Compiegne}, {Davis}, {de
  Bernardis}, {de Gasperis}, {Di Francesco}, {di Giorgio}, {Elia}, {Faustini},
  {Fischera}, {Fukui}, {Fuller}, {Ganga}, {Garcia-Lario}, {Giard}, {Giardino},
  {Glenn}, {Goldsmith}, {Griffin}, {Hoare}, {Huang}, {Jiang}, {Joblin},
  {Joncas}, {Juvela}, {Kirk}, {Lagache}, {Li}, {Lim}, {Lord}, {Lucas},
  {Maiolo}, {Marengo}, {Marshall}, {Masi}, {Massi}, {Matsuura}, {Meny},
  {Minier}, {Miville-Desch{\^e}nes}, {Montier}, {Motte}, {M{\"u}ller},
  {Natoli}, {Neves}, {Olmi}, {Paladini}, {Paradis}, {Pestalozzi}, {Pezzuto},
  {Piacentini}, {Pomar{\`e}s}, {Popescu}, {Reach}, {Richer}, {Ristorcelli},
  {Roy}, {Royer}, {Russeil}, {Saraceno}, {Sauvage}, {Schilke},
  {Schneider-Bontemps}, {Schuller}, {Schultz}, {Shepherd}, {Sibthorpe},
  {Smith}, {Smith}, {Spinoglio}, {Stamatellos}, {Strafella}, {Stringfellow},
  {Sturm}, {Taylor}, {Thompson}, {Tuffs}, {Umana}, {Valenziano}, {Vavrek},
  {Viti}, {Waelkens}, {Ward-Thompson}, {White}, {Wyrowski}, {Yorke}, \&
  {Zhang}}]{mol10a}
{Molinari}, S., {Swinyard}, B., {Bally}, J., {et~al.} 2010{\natexlab{a}},
  \pasp, 122, 314

\bibitem[{{Molinari} {et~al.}(2010{\natexlab{b}}){Molinari}, {Swinyard},
  {Bally}, {Barlow}, {Bernard}, {Martin}, {Moore}, {Noriega-Crespo}, {Plume},
  {Testi}, {Zavagno}, {Abergel}, {Ali}, {Anderson}, {Andr{\'e}}, {Baluteau},
  {Battersby}, {Beltr{\'a}n}, {Benedettini}, {Billot}, {Blommaert}, {Bontemps},
  {Boulanger}, {Brand}, {Brunt}, {Burton}, {Calzoletti}, {Carey}, {Caselli},
  {Cesaroni}, {Cernicharo}, {Chakrabarti}, {Chrysostomou}, {Cohen},
  {Compiegne}, {de Bernardis}, {de Gasperis}, {di Giorgio}, {Elia}, {Faustini},
  {Flagey}, {Fukui}, {Fuller}, {Ganga}, {Garcia-Lario}, {Glenn}, {Goldsmith},
  {Griffin}, {Hoare}, {Huang}, {Ikhenaode}, {Joblin}, {Joncas}, {Juvela},
  {Kirk}, {Lagache}, {Li}, {Lim}, {Lord}, {Marengo}, {Marshall}, {Masi},
  {Massi}, {Matsuura}, {Minier}, {Miville-Deschenes}, {Montier}, {Morgan},
  {Motte}, {Mottram}, {Mueller}, {Natoli}, {Neves}, {Olmi}, {Paladini},
  {Paradis}, {Parsons}, {Peretto}, {Pestalozzi}, {Pezzuto}, {Piacentini},
  {Piazzo}, {Polychroni}, {Pomar{\`e}s}, {Popescu}, {Reach}, {Ristorcelli},
  {Robitaille}, {Robitaille}, {Rod{\'o}n}, {Roy}, {Royer}, {Russeil},
  {Saraceno}, {Sauvage}, {Schilke}, {Schisano}, {Schneider}, {Schuller},
  {Schulz}, {Sibthorpe}, {Smith}, {Smith}, {Spinoglio}, {Stamatellos},
  {Strafella}, {Stringfellow}, {Sturm}, {Taylor}, {Thompson}, {Traficante},
  {Tuffs}, {Umana}, {Valenziano}, {Vavrek}, {Veneziani}, {Viti}, {Waelkens},
  {Ward-Thompson}, {White}, {Wilcock}, {Wyrowski}, {Yorke}, \&
  {Zhang}}]{mol10b}
{Molinari}, S., {Swinyard}, B., {Bally}, J., {et~al.} 2010{\natexlab{b}}, ArXiv
  e-prints

\bibitem[{{Mottram} {et~al.}(2010){Mottram}, {Hoare}, {Lumsden}, {Oudmaijer},
  {Urquhart}, {Meade}, {Moore}, \& {Stead}}]{mot10}
{Mottram}, J.~C., {Hoare}, M.~G., {Lumsden}, S.~L., {et~al.} 2010, \aap, 510,
  A89+

\bibitem[{{Omont} {et~al.}(2003){Omont}, {Gilmore}, {Alard}, {Aracil},
  {August}, {Baliyan}, {Beaulieu}, {B{\'e}gon}, {Bertou}, {Blommaert},
  {Borsenberger}, {Burgdorf}, {Caillaud}, {Cesarsky}, {Chitre}, {Copet}, {de
  Batz}, {Egan}, {Egret}, {Epchtein}, {Felli}, {Fouqu{\'e}}, {Ganesh},
  {Genzel}, {Glass}, {Gredel}, {Groenewegen}, {Guglielmo}, {Habing},
  {Hennebelle}, {Jiang}, {Joshi}, {Kimeswenger}, {Messineo},
  {Miville-Desch{\^e}nes}, {Moneti}, {Morris}, {Ojha}, {Ortiz}, {Ott},
  {Parthasarathy}, {P{\'e}rault}, {Price}, {Robin}, {Schultheis}, {Schuller},
  {Simon}, {Soive}, {Testi}, {Teyssier}, {Tiph{\`e}ne}, {Unavane}, {van Loon},
  \& {Wyse}}]{omo03}
{Omont}, A., {Gilmore}, G.~F., {Alard}, C., {et~al.} 2003, \aap, 403, 975

\bibitem[{{Ossenkopf} \& {Henning}(1994)}]{oss94}
{Ossenkopf}, V. \& {Henning}, T. 1994, \aap, 291, 943

\bibitem[{{Ott}(2010)}]{ott10}
{Ott}, S. 2010, ASP Conference Series, Astronomical Data Analysis Software and
  System XIX, in press

\bibitem[{{Paradis} {et~al.}(2010){Paradis}, {Veneziani}, {Noriega-Crespo},
  {Paladini}, {Piacentini}, {Bernard}, {de Bernardis}, {Calzoletti},
  {Faustini}, {Martin}, {Masi}, {Montier}, {Natoli}, {Ristorcelli}, {Thompson},
  {Traficante}, \& {Molinari}}]{par10}
{Paradis}, D., {Veneziani}, M., {Noriega-Crespo}, A., {et~al.} 2010, \aap, 520,
  L8+

\bibitem[{{Parsons} {et~al.}(2009){Parsons}, {Thompson}, \&
  {Chrysostomou}}]{par09}
{Parsons}, H., {Thompson}, M.~A., \& {Chrysostomou}, A. 2009, \mnras, 399, 1506

\bibitem[{{Perault} {et~al.}(1996){Perault}, {Omont}, {Simon}, {Seguin},
  {Ojha}, {Blommaert}, {Felli}, {Gilmore}, {Guglielmo}, {Habing}, {Price},
  {Robin}, {de Batz}, {Cesarsky}, {Elbaz}, {Epchtein}, {Fouque}, {Guest},
  {Levine}, {Pollock}, {Prusti}, {Siebenmorgen}, {Testi}, \& {Tiphene}}]{per96}
{Perault}, M., {Omont}, A., {Simon}, G., {et~al.} 1996, \aap, 315, L165

\bibitem[{{Peretto} \& {Fuller}(2009)}]{per09}
{Peretto}, N. \& {Fuller}, G.~A. 2009, \aap, 505, 405

\bibitem[{{Peretto} \& {Fuller}(2010)}]{per10b}
{Peretto}, N. \& {Fuller}, G.~A. 2010, \apj, 723, 555

\bibitem[{{Peretto} {et~al.}(2010){Peretto}, {Fuller}, {Plume}, {Anderson},
  {Bally}, {Battersby}, {Beltran}, {Bernard}, {Calzoletti}, {Digiorgio},
  {Faustini}, {Kirk}, {Lenfestey}, {Marshall}, {Martin}, {Molinari}, {Montier},
  {Motte}, {Ristorcelli}, {Rod{\'o}n}, {Smith}, {Traficante}, {Veneziani},
  {Ward-Thompson}, \& {Wilcock}}]{per10}
{Peretto}, N., {Fuller}, G.~A., {Plume}, R., {et~al.} 2010, \aap, 518, L98+

\bibitem[{{Pestalozzi} {et~al.}(2005){Pestalozzi}, {Minier}, \&
  {Booth}}]{pes05}
{Pestalozzi}, M.~R., {Minier}, V., \& {Booth}, R.~S. 2005, \aap, 432, 737

\bibitem[{{Pilbratt} {et~al.}(2010){Pilbratt}, {Riedinger}, {Passvogel},
  {Crone}, {Doyle}, {Gageur}, {Heras}, {Jewell}, {Metcalfe}, {Ott}, \&
  {Schmidt}}]{pil10}
{Pilbratt}, G.~L., {Riedinger}, J.~R., {Passvogel}, T., {et~al.} 2010, ArXiv
  e-prints

\bibitem[{{Poglitsch} {et~al.}(2010){Poglitsch}, {Waelkens}, {Geis},
  {Feuchtgruber}, {Vandenbussche}, {Rodriguez}, {Krause}, {Renotte}, {van
  Hoof}, {Saraceno}, {Cepa}, {Kerschbaum}, {Agnese}, {Ali}, {Altieri},
  {Andreani}, {Augueres}, {Balog}, {Barl}, {Bauer}, {Belbachir}, {Benedettini},
  {Billot}, {Boulade}, {Bischof}, {Blommaert}, {Callut}, {Cara}, {Cerulli},
  {Cesarsky}, {Contursi}, {Creten}, {De Meester}, {Doublier}, {Doumayrou},
  {Duband}, {Exter}, {Genzel}, {Gillis}, {Gr{\"o}zinger}, {Henning},
  {Herreros}, {Huygen}, {Inguscio}, {Jakob}, {Jamar}, {Jean}, {de Jong},
  {Katterloher}, {Kiss}, {Klaas}, {Lemke}, {Lutz}, {Madden}, {Marquet},
  {Martignac}, {Mazy}, {Merken}, {Montfort}, {Morbidelli}, {M{\"u}ller},
  {Nielbock}, {Okumura}, {Orfei}, {Ottensamer}, {Pezzuto}, {Popesso},
  {Putzeys}, {Regibo}, {Reveret}, {Royer}, {Sauvage}, {Schreiber}, {Stegmaier},
  {Schmitt}, {Schubert}, {Sturm}, {Thiel}, {Tofani}, {Vavrek}, {Wetzstein},
  {Wieprecht}, \& {Wiezorrek}}]{pog10}
{Poglitsch}, A., {Waelkens}, C., {Geis}, N., {et~al.} 2010, ArXiv e-prints

\bibitem[{{Ragan} {et~al.}(2006){Ragan}, {Bergin}, {Plume}, {Gibson}, {Wilner},
  {O'Brien}, \& {Hails}}]{rag06}
{Ragan}, S.~E., {Bergin}, E.~A., {Plume}, R., {et~al.} 2006, \apjs, 166, 567

\bibitem[{{Rathborne} {et~al.}(2006){Rathborne}, {Jackson}, \& {Simon}}]{rat06}
{Rathborne}, J.~M., {Jackson}, J.~M., \& {Simon}, R. 2006, \apj, 641, 389

\bibitem[{{Reid} {et~al.}(2009){Reid}, {Menten}, {Zheng}, {Brunthaler},
  {Moscadelli}, {Xu}, {Zhang}, {Sato}, {Honma}, {Hirota}, {Hachisuka}, {Choi},
  {Moellenbrock}, \& {Bartkiewicz}}]{rei09}
{Reid}, M.~J., {Menten}, K.~M., {Zheng}, X.~W., {et~al.} 2009, ArXiv e-prints

\bibitem[{{Rosolowsky} {et~al.}(2010){Rosolowsky}, {Dunham}, {Ginsburg},
  {Bradley}, {Aguirre}, {Bally}, {Battersby}, {Cyganowski}, {Dowell},
  {Drosback}, {Evans}, {Glenn}, {Harvey}, {Stringfellow}, {Walawender}, \&
  {Williams}}]{ros10}
{Rosolowsky}, E., {Dunham}, M.~K., {Ginsburg}, A., {et~al.} 2010, \apjs, 188,
  123

\bibitem[{{Schuller} {et~al.}(2009){Schuller}, {Menten}, {Contreras},
  {Wyrowski}, {Schilke}, {Bronfman}, {Henning}, {Walmsley}, {Beuther},
  {Bontemps}, {Cesaroni}, {Deharveng}, {Garay}, {Herpin}, {Lefloch}, {Linz},
  {Mardones}, {Minier}, {Molinari}, {Motte}, {Nyman}, {Reveret}, {Risacher},
  {Russeil}, {Schneider}, {Testi}, {Troost}, {Vasyunina}, {Wienen}, {Zavagno},
  {Kovacs}, {Kreysa}, {Siringo}, \& {Wei{\ss}}}]{sch09}
{Schuller}, F., {Menten}, K.~M., {Contreras}, Y., {et~al.} 2009, \aap, 504, 415

\bibitem[{{Simon} {et~al.}(2006){Simon}, {Jackson}, {Rathborne}, \&
  {Chambers}}]{sim06}
{Simon}, R., {Jackson}, J.~M., {Rathborne}, J.~M., \& {Chambers}, E.~T. 2006,
  \apj, 639, 227

\bibitem[{{Stil} {et~al.}(2006){Stil}, {Taylor}, {Dickey}, {Kavars}, {Martin},
  {Rothwell}, {Boothroyd}, {Lockman}, \& {McClure-Griffiths}}]{sti06}
{Stil}, J.~M., {Taylor}, A.~R., {Dickey}, J.~M., {et~al.} 2006, \aj, 132, 1158

\bibitem[{{Szymczak} {et~al.}(2002){Szymczak}, {Kus}, {Hrynek}, {K{\v e}pa}, \&
  {Pazderski}}]{szy02}
{Szymczak}, M., {Kus}, A.~J., {Hrynek}, G., {K{\v e}pa}, A., \& {Pazderski}, E.
  2002, \aap, 392, 277

\bibitem[{{Traficante} {et~al.}(2010){Traficante}, {Calzoletti}, \&
  {Veneziani}}]{tra10}
{Traficante}, A., {Calzoletti}, L., \& {Veneziani}, M. 2010, submitted

\bibitem[{{Walsh} {et~al.}(1998){Walsh}, {Burton}, {Hyland}, \&
  {Robinson}}]{wal98}
{Walsh}, A.~J., {Burton}, M.~G., {Hyland}, A.~R., \& {Robinson}, G. 1998,
  \mnras, 301, 640

\bibitem[{{White} {et~al.}(2005){White}, {Becker}, \& {Helfand}}]{whi05}
{White}, R.~L., {Becker}, R.~H., \& {Helfand}, D.~J. 2005, \aj, 130, 586

\end{thebibliography}

\end{document}